\documentclass{jfm} 
\pdfoutput=1

\usepackage{amsmath}
\usepackage{graphicx}
\usepackage{dcolumn}
\usepackage{bm}
\usepackage{hyperref}
\usepackage{lineno}
\usepackage{color}

\usepackage{tikz}
\usepackage{pgfplots}
\usetikzlibrary{shapes}

\usepackage{amssymb}
\usepackage{appendix}

\newcommand{\cb}{\color{black}}

\def\XXint#1#2#3{{\setbox0=\hbox{$#1{#2#3}{\int}$}
     \vcenter{\hbox{$#2#3$}}\kern-.5\wd0}}

\usepackage{fancyhdr} 
\shorttitle{SSP theory for UMZs and internal layers}
\shortauthor{B. Montemuro and C. M. White and J. Klewicki and G. P. Chini}

\title{A self-sustaining process theory for uniform momentum zones and internal shear layers in high Reynolds number shear flows}

\author{Brandon Montemuro\aff{1},
  Christopher M. White\aff{2},
  Joseph C. Klewicki\aff{1,2,3},
  \and Gregory P. Chini\aff{1,2}\corresp{\email{greg.chini@unh.edu}}
  }

\affiliation{\aff{1} Program in Integrated Applied Mathematics, University of New Hampshire, Durham, NH 03824, USA
\aff{2} Department of Mechanical Engineering, University of New Hampshire, Durham, NH 03824, USA
\aff{3} Department of Mechanical Engineering, University of Melbourne, Melbourne, Victoria 3010, Australia}

\begin{document}
\maketitle
\begin{abstract}
Many exact coherent states (ECS) arising in wall-bounded shear flows have an asymptotic structure at extreme Reynolds number $Re$ in which the \emph{effective} Reynolds number governing the streak and roll dynamics is $\mathit{O}(1)$.  Consequently, these viscous ECS are not suitable candidates for quasi-coherent structures away from the wall that necessarily are inviscid in the mean.
Specifically, viscous ECS cannot account for the singular nature of the inertial domain, where the flow self-organizes into uniform momentum zones (UMZs) separated by internal shear layers and the instantaneous streamwise velocity develops a staircase-like profile.  In this investigation, a large-$Re$ asymptotic analysis is performed to explore the potential for a three-dimensional, short streamwise- and spanwise-wavelength instability of the embedded shear layers to sustain a spatially-distributed array of much larger-scale, effectively inviscid streamwise roll motions.  In contrast to other self-sustaining process theories, the rolls are sufficiently strong to differentially homogenize the background shear flow, thereby providing a mechanistic explanation for the formation and maintenance of UMZs and interlaced shear layers that respects the leading-order balance structure of the mean dynamics.
\end{abstract}
\section{Introduction}\label{sec:INTRO}
\begin{figure}
	\begin{center}
             \includegraphics[width=0.99\linewidth]{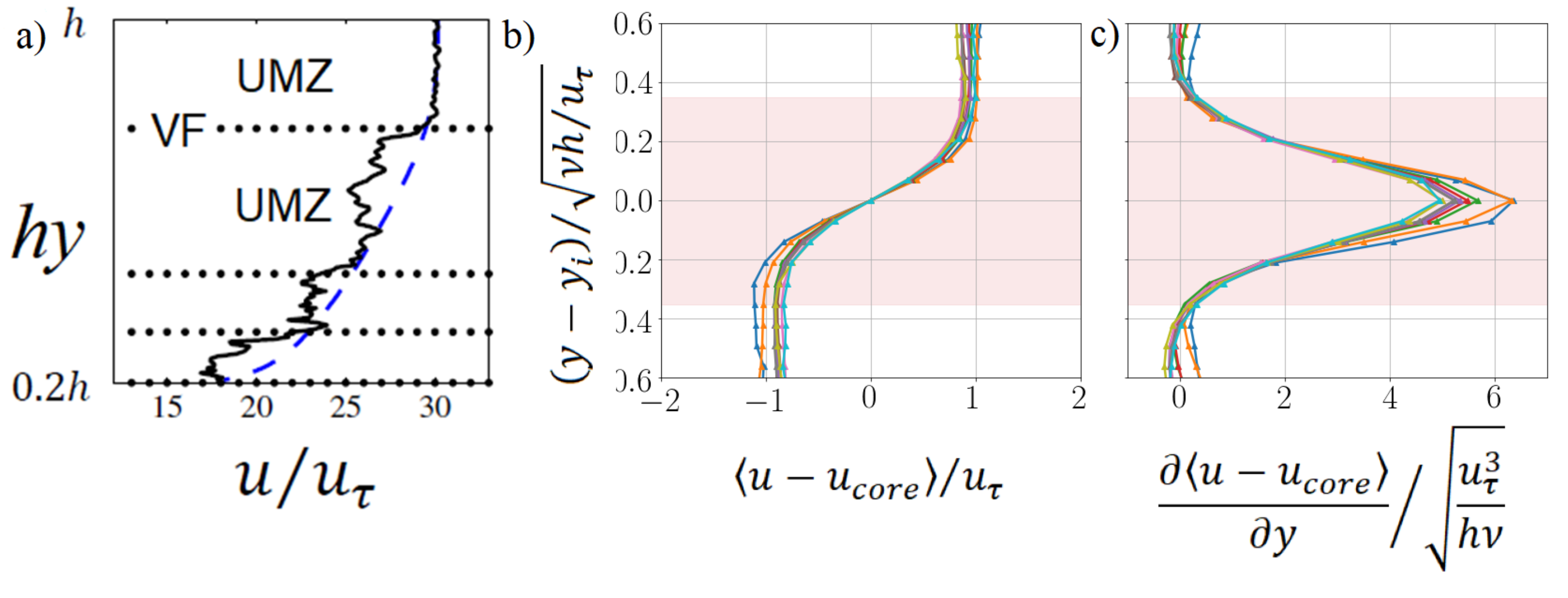}
	\caption{Uniform momentum zones (UMZs) and internal shear layers (or \emph{vortical fissures}, VFs) are ubiquitous in the inertial region of turbulent wall flows at sufficiently large values of the friction Reynolds number $Re_\tau$.  (a)~Wall-normal profile of {\cb{instantaneous}} streamwise velocity taken from the boundary layer (BL) measurements of \cite{deSilvaJFM2016} at $Re_\tau\approx 8000$ illustrating the staircase-like arrangement of UMZs and VFs.  (b,c)~ Conditionally-averaged streamwise velocity (b) and its wall-normal derivative (c) through vortical fissures in the inertial region of a turbulent BL flow (at $Re_\tau\approx 6000$) in the UNH FPF. $y_i$ indicates the wall-normal location of the VF; $u_{core}$ is the streamwise velocity at the center of the VF; and angle brackets indicate a conditional average, {\cb{where the conditioning is based on a spanwise vorticity threshold equal to $3\sqrt{Re_\tau}(u_\tau/h)$.  The various distinct conditionally-averaged profiles are obtained by segregating the instantaneous profiles into 10 contiguous wall-normal bins (all located within the inertial domain) spanning the measurement field of view.}}}
	\label{fig:conditional_u_omega}
	\end{center}
\end{figure}

Beginning with the seminal work of \cite{MeinhartAdrianPoF1995}, an increasing number of laboratory \citep{deSilvaJFM2016,deSilvaJFM2017}, field 
\citep{PriyaJFM2007} and computational studies \citep{LeeMoserJFM2015, LeeMoserJFM2018, HickeyJFM2019} has revealed the remarkable tendency for turbulent wall flows to self-organize into regions of quasi-uniform streamwise momentum segregated by internal layers of concentrated spanwise vorticity.  As shown in figure~\ref{fig:conditional_u_omega}(a), taken from the experiments of \cite{deSilvaJFM2016}, the resulting arrangement of uniform momentum zones (UMZs) and internal shear layers, which we refer to as \emph{vortical fissures} (VFs), causes the {\cb{instantaneous}} wall-normal ($y$) profile of streamwise ($x)$ velocity to exhibit a staircase-like structure.  The largest UMZs remain coherent in the streamwise direction for distances approaching the boundary layer (BL) thickness or channel half-height $h$ for times estimated to be {\cb{$\mathit{O}(h/U_\infty)$, where $U_\infty$ is the free-stream or mean channel-centerline velocity \citep{LaskariJFM2018}.}}
Both laboratory experiments \citep{deSilvaJFM2016} and analyses of the mean momentum equation \citep{KlewickiJFM2013,KlewickiIUTAM2013} indicate that the number of UMZs grows logarithmically with the friction Reynolds number $Re_\tau\equiv u_\tau h/\nu$, {\cb{where $u_\tau$ is the wall friction velocity and}} $\nu$ is the kinematic viscosity of the fluid, and that the typical change in streamwise velocity across individual fissures is a few multiples of $u_\tau$.  
{\cb{The aim of the present investigation is to propose a mechanistic explanation for the emergence of UMZs and VFs by deriving directly from the governing Navier--Stokes equations a new, multiscale self-sustaining process (SSP) that supports exact coherent states (ECS) exhibiting these two flow features.}}

This UMZ/VF structure is most readily discerned in conditionally-averaged profiles of the streamwise velocity, with the conditioning generally being based on a spanwise ($z$) vorticity threshold or a discrete approximation to the probability density function of streamwise velocity. For example, figure~\ref{fig:conditional_u_omega}(b,c) shows the conditionally-averaged local profiles of streamwise velocity (b) and spanwise vorticity (c) through VFs detected in the inertial region of a high Reynolds number turbulent BL in the Flow Physics Facility (FPF) at the University of New Hampshire (UNH).  The hyperbolic-tangent-like shape of the streamwise velocity profile is striking and robustly observed in these conditional averages.   These results also confirm that the characteristic jump in speed across the internal shear layer (i.e. the VF) separating adjacent UMZs is $\mathit{O}(u_\tau)$.

As evident in figure~\ref{fig:conditional_u_omega}(c), 
the conditionally-averaged spanwise vorticity
within the VF is appropriately scaled by $\sqrt{u_\tau^{3}/(h \nu)}$.  This scaling suggests that the typical thickness of a VF is proportional to $h/\sqrt{Re_\tau}$ and therefore decreases (in outer units) as $Re_\tau$ increases.  Of course, the data used to generate the vorticity profile in figure~\ref{fig:conditional_u_omega} were obtained at a fixed Reynolds number ($Re_\tau\approx 6000$).  Nevertheless, other laboratory and field measurements similarly indicate that the dimensional VF thickness $\Delta_f$ normalized by the BL height $h$ scales in proportion to $1/\sqrt{Re_\tau}$ over the range $10^3<Re_\tau<10^6$ \citep{KlewickiJFM2013} and very likely for higher $Re_\tau$; see figure~\ref{fig:FissureThickness}(a) in \S~\ref{sec:LIMITATION}.  Given that the jump in flow speed across each VF appears to be independent of $Re_\tau$ \citep{deSilvaJFM2017}, a crucial implication of this scaling is that an increasing percentage of the total spanwise vorticity is confined to an ever diminishing fraction of the wall-normal flow domain as $Re_\tau\to\infty$.  

\begin{figure}
	\begin{center}
	\includegraphics[width=0.95\linewidth]{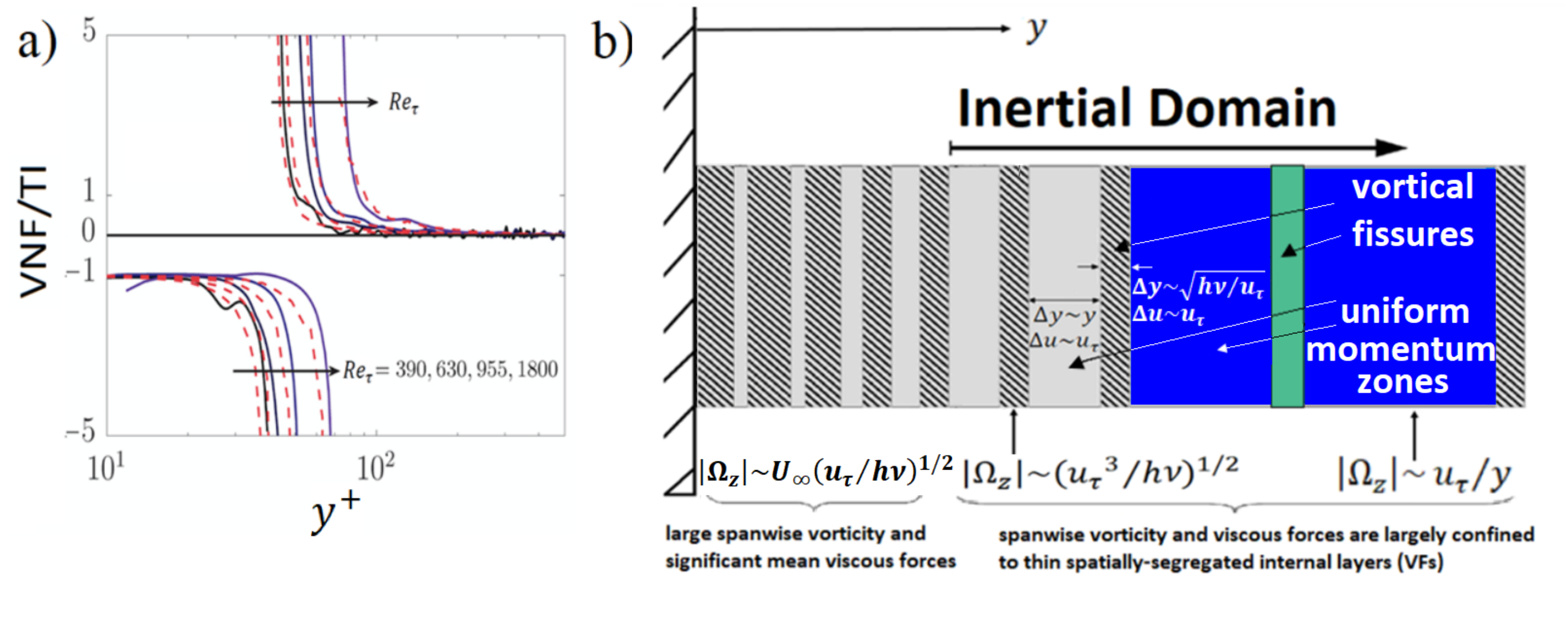}
	\caption{The singular nature of the turbulent boundary layer.  (a)~Plot showing the ratio of the {\cb{mean viscous normal force (VNF) to turbulent inertia (TI),}} i.e. to the Reynolds stress gradient, in turbulent channel flow at four different values of $Re_\tau$; data taken from the water channel experiments by \cite{ElsnabEXPFLUIDS2017}.  Note that the mean viscous force is significant in a volume-averaged sense only in a near-wall domain of size $\mathit{O}(\sqrt{Re_\tau})$ in viscous (or ``plus") units, corresponding to a domain of size $\mathit{O}(h/\sqrt{Re_\tau})$ in outer units.  Outboard of the peak in the Reynolds stress (where the force ratio tends to plus or minus infinity), the volume-averaged mean viscous force is negligible.  (b)~Schematic illustrating the concentration of spanwise vorticity within VFs that, at large $Re_\tau$, become increasingly widely separated with increasing distance from the wall. {\cb{(Adapted from \cite{KlewickiJFM2013,KlewickiIUTAM2013}.) The new SSP theory developed herein targets UMZs and VFs located in the inertial domain, as highlighted in blue and green, respectively.}}}
\label{fig:mean_force_balance}
\end{center}
\end{figure}

This striking behavior is one manifestation of the absence of leading-order mean viscous forces on the domain of interest.  
Indeed, the UMZ/VF profiles of streamwise velocity are most prominent within the \emph{inertial region} of turbulent wall flows, the region in which the mean viscous force is negligible relative to turbulent inertia (i.e. to the Reynolds stress gradient) and to mean inertia and/or the mean pressure gradient; see figure~\ref{fig:mean_force_balance}(a).  Using a complement of theory and corroborating empiricism, \cite{WeiJFM2005} demonstrate that the inertial domain emerges at a distance $\mathit{O}(h/\sqrt{Re_\tau})$ from the wall and spans a region that is $\mathit{O}(h)$ in wall-normal extent.  
{\cb{Informed by these results and by similarity analysis of the mean momentum equation, \cite{KlewickiJFM2013,KlewickiIUTAM2013} proposed}} that the turbulent boundary layer is singular as $Re_\tau\to\infty$ in two complementary ways.  Firstly, there is a region of thickness $\mathit{O}(h/\sqrt{Re_\tau})$ directly adjacent to the wall in which mean viscous forces and spanwise vorticity are significant.  Unlike their laminar counterparts, however, turbulent BLs also exhibit regions of intense spanwise vorticity away from the wall -- but only in asymptotically thin, spatially-segregated domains, i.e. within VFs.  The emerging paradigm, therefore, is that at extreme values of the Reynolds number the turbulent boundary layer comprises logarithmically many viscous (albeit not necessarily laminar) internal layers of elevated vorticity that are spatially dispersed throughout the bulk of the BL volume (figure~\ref{fig:mean_force_balance}b).

To quantitatively assess this paradigm, \cite{BautistaJFM2018}, in companion work, recently developed a one-dimensional (1D) model of turbulent wall flows that exploits the increasingly binary spatial distribution of the spanwise vorticity field at large $Re_\tau$.  A master staircase-like profile of the instantaneous streamwise velocity is constructed by incorporating just two elements, VFs and UMZs, with the wall-normal locations of the VFs and the associated increments in streamwise flow speed specified in accord with the similarity reduction of the mean momentum equation performed by \cite{KlewickiJFM2013}.  To generate instantaneous realizations, the fissures are randomly displaced (using an empirical positively-skewed Gaussian distribution), exchanging momentum as they are redistributed.   Flow statistics are obtained by ensemble averaging sufficiently many realizations generated according to this protocol.  The resulting mean profiles of streamwise velocity, velocity variance, and sub-Gaussian skewness and kurtosis agree remarkably well with those acquired from direct numerical simluations (DNS) of turbulent channel flow at large $Re_\tau$, lending considerable credence to the ``boundary-layers-within-\emph{the}-turbulent-boundary-layer" paradigm.

\begin{figure}
	\begin{center}
		\includegraphics[width=1.0\linewidth]{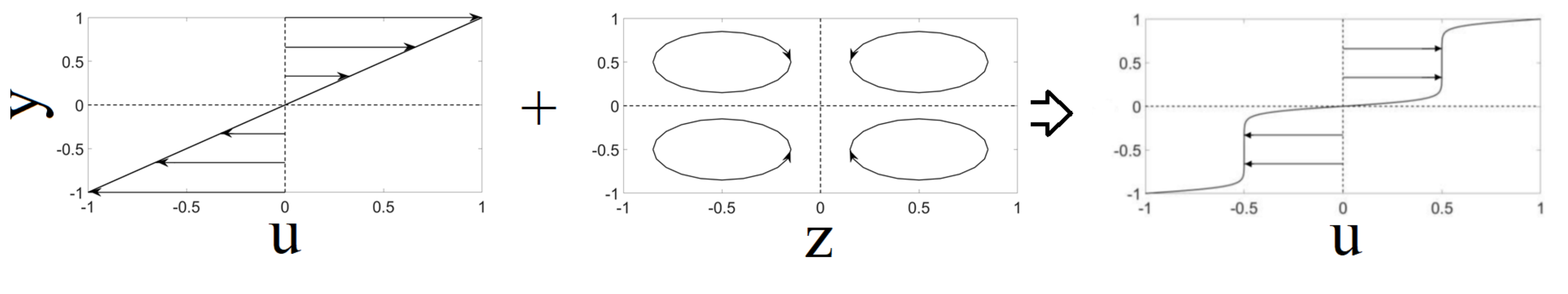}
		\caption{Proposed flow configuration in which sufficiently strong {\cb{counter-rotating rolls, stacked in the $y$-$z$ plane,}} redistribute the imposed background shear in the streamwise velocity (here taken to be unbounded Couette flow) yielding a staircase-like UMZ/VF profile.}
		\label{fig:Proposed}
	\end{center}
\end{figure}

For all its merit, the 1D turbulence model developed by \cite{BautistaJFM2018} cannot address a fundamental mechanistic question; namely, why should regions of quasi-uniform momentum with embedded shear layers spontaneously arise in an otherwise smoothly sheared flow?  
Indeed, despite their probable physical importance to turbulent transport, the origin of UMZs and VFs is not clearly understood.  For example,  \cite{KimAdrianPoF1999} implicate large-scale and very-large-scale motions (LSMs and VLSMs), arguing that these large-scale flow structures arise from the spontaneous organization of smaller-scale attached and/or detached hairpin vortices and hairpin vortex packets.  Employing a resolvent-based model for LSMs, \cite{McKeonJFMRapids2017} find evidence of UMZ signatures, although the Reynolds number scaling of the VFs (if any) is not addressed.  Hwang and collaborators \citep{HwangCossuPRL2010,RawatJFM2015,HwangJFMrapids2016,HwangJFM2016} use a modified large-eddy simulation (LES) scheme to demonstrate that LSMs and VLSMs can be \emph{directly} sustained -- {\cb{even in the absence of small-scale near-wall structures -- via a ``filtered'' self-sustaining process}}, but they do not investigate the formation of UMZs or VFs.  Like these latter authors, our premise is that inertial-region structures are supported directly through a {\cb{large-$Re_\tau$}} SSP, but our focus is on the sustenance of UMZs and interlaced VFs, and our first-principles approach is grounded in the instantaneous rather than LES-filtered Navier--Stokes (NS) equations. 

Building on our earlier theoretical investigation \citep{ChiniPTRSA2017}, we hypothesize that sufficiently strong large-scale counter-rotating streamwise rolls, having a diameter much greater than the VF thickness and stacked in the wall-normal direction, can differentially homogenize an imposed background shear flow. {\cb{As shown in figure~\ref{fig:Proposed}, the result is a pattern of UMZs and embedded internal shear layers (i.e. VFs). Streamwise rolls and the streaks they induce are key components of equilibrium, traveling-wave and periodic-orbit exact coherent states arising in incompressible wall-bounded shear flows \citep{WaleffePoF1997, FaistPRL2003, WedinJFM2004, NagataJFM1990, DuguetJFM2008, GibsonJFM2008, DuguetJFM2010}. ECS necessarily are self-sustaining since, by construction, they are \emph{invariant} solutions of the NS equations.  Accordingly, we seek a mechanistic explanation for the occurrence of VFs and UMZs by deriving from the NS equations an asymptotic SSP formalism whose ECS solutions exhibit these flow features.}}

{\cb{To date, there are arguably two distinct first-principles SSP theories that can account for ECS in constant-density wall-bounded parallel shear flows. Most germane to the present investigation is the classical self-sustaining process theory introduced by \cite{WaleffePoF1997} and the closely related asymptotic vortex--wave interaction (VWI) formalism developed earlier by \cite{HallSmithJFM1991} and subsequently applied to ECS in wall-bounded shear flows by \cite{HallSherwinJFM2010}.  In both Waleffe's SSP and in VWI theory (the latter may be viewed as the infinite Reynolds number limit of the former), the nonlinear self-interaction of a streamwise-varying instability ``wave'' drives the roll motions that advect the base shear flow to generate streaks.  Since the streak profile is inflectional, an inviscid Rayleigh instability mode is excited. (Alternatively, in channel flows, a near-wall Tollmien--Schlichting wave can instead be implicated in the SSP, as shown by \cite{DempseyJFM2016}.)   At large Reynolds number, the nonlinear self-interaction of the Rayleigh mode is confined to a critical layer where the ECS phase speed matches the streak velocity. \cite{BlackburnHallSherwinJFM2013} and \cite{EckhardtZammertNONLINEARITY2018} have demonstrated that VWI states can exist on ever smaller spatial scales as the Reynolds number is increased. Specifically, when the streamwise and spanwise wavenumbers are increased such that their ratio remains fixed, the ECS adopt a self-similar form in which the coherent structure becomes localised in the wall-normal direction within a ``production'' layer having a thickness comparable to the inverse spanwise (or streamwise) wavenumber.  \cite{DeguchiJFM2015} used scaling analysis and full NS computations to show that when the wavenumbers become sufficiently large, convergence to VWI states is lost. Instead, a new class of ECS emerges, supported by a distinct SSP in which the velocity deviation from the imposed laminar shear flow scales in proportion to the inverse Reynolds number and consequently satisfies the full NS equations but at unit Reynolds number.  A similar ``unit-Reynolds-number NS'' (UNS) mechanism underlies the free-stream ECS identified by \cite{DeguchiHallJFM2014a}, except that a large-amplitude, passive streak is driven outside the production layer.  Note that for UNS ECS, a decomposition into wave, roll and streak flow components is not particularly meaningful, since all ECS velocities scale in proportion to the inverse Reynolds number.}}


{\cb{In the present theory, as in VWI analysis (see \S~\ref{sec:LIMITATION}), the streamwise-invariant rolls are sustained by the nonlinear self-interaction of a streamwise-varying instability (here termed a ``fluctuation") mode.  Unlike VWI theory, however, the Rayleigh instability mode supported by the wall-normal inflection in the streamwise velocity}} has a streamwise wavelength commensurate with the thickness of the embedded shear layer; that is, the streamwise wavelength of the fluctuation is small compared to the scale of the rolls, a crucial refinement of our prior theory \citep{ChiniPTRSA2017}.  Moreover, as demonstrated in \S~\ref{sec:ASYMPTOTICS}, the instability mode is refracted and rendered three-dimensional (3D) owing to the comparably slow spanwise variation in the thickness of the embedded VF that is caused by the roll-scale spanwise variation of the streak velocity.  {\cb{(The target structure is displayed in figure~\ref{fig:Reconstruction} in \S~\ref{sec:ECS}.)}} 

The primary objective of the present theoretical investigation is to demonstrate that the proposed multiscale SSP is, in fact, compatible with the governing incompressible NS equations in the limit of asymptotically large Reynolds number.  As described in \S~\ref{sec:ASYMPTOTICS}, we do this by performing a multiscale asymptotic analysis of unbounded plane Couette flow, an imperfect but nevertheless useful surrogate for the inertial domain of turbulent wall flows.  New, spatially-distributed ECS supported by this SSP are documented in \S~\ref{sec:ECS}.  In \S~\ref{sec:CONCLUSION}, we discuss the applicability of our SSP theory to the inertial domain of wall-bounded turbulent shear flows and suggest avenues for future investigation. Before describing the asymptotic analysis, we first (in \S~\ref{sec:LIMITATION}) draw an important distinction between our proposed SSP and {\cb{VWI theory}}.

\section{Viscous versus inertial exact coherent states}\label{sec:LIMITATION}

{\cb{Invariant}} solutions of the NS equations are believed to provide a scaffold in phase space for the turbulent dynamics realized at large Reynolds number \citep{GibsonJFM2008, GibsonJFM2009, SuriPRL2017}.  Nevertheless, many ECS -- including upper-branch states -- have an asymptotic structure in which the \emph{effective} Reynolds number governing the streak and roll dynamics is order unity.  Although these viscous, space-filling ECS likely play a role in the dynamics of the near-wall region, they cannot be relevant to the inertial layer (outboard of the Reynolds stress peak), where the leading-order mean dynamics is known to be inviscid (figure~\ref{fig:mean_force_balance}a).  In particular, viscous ECS cannot account for the uniform momentum zones and internal shear layers observed in the inertial domain.  

To support this assertion, we first decompose the velocity field into a streamwise- or $x$-averaged component, denoted with an overbar, plus an $x$-dependent fluctuation field, denoted with a prime.  Considering, for simplicity, plane Couette flow, for which the mean streamwise pressure gradient vanishes, the dimensionless $x$-averaged streamwise momentum equation reduces to  
\begin{eqnarray}
\partial_t\bar{u}\,+\,\bar{v}\partial_y\bar{u}\,+\,\bar{w}\partial_z\bar{u}&=&\frac{1}{Re}\left(\partial_y^2+\partial_z^2\right)\bar{u}\,+\,\mbox{H.O.T.}\label{eq:ADVDIFF}
\end{eqnarray}
{\cb{Here, $Re$ is a Reynolds number defined based on the speed of the channel wall and the channel half-height,}} and $u$, $v$ and $w$ are the (dimensionless) velocity components in the $x$, $y$ and $z$ directions, respectively.  H.O.T. refers to nonlinear correlations that are small compared to the terms retained provided that the $x$-varying fluctuations $(u',v',w')\ll\bar{u}$ as $Re\to\infty$.  This proviso holds for Waleffe's SSP  \citep{WangPRL2007} and for the VWI theory of Hall and Smith \citep{HallSmithJFM1991,HallSherwinJFM2010}.  Crucially, in both theories, the streamwise roll motions, represented by the mean advection velocity ($\bar{v}$,$\bar{w}$) in the $y$--$z$ plane, are $\mathit{O}(1/Re)$ and, consequently, the effective Reynolds number governing the dynamics of the $\mathit{O}(1)$ streak\footnote{Strictly, \emph{streak} refers to the difference between the streamwise- and horizontally-averaged streamwise flow: $\bar{u}-\overline{u}^{xz}$, where $\overline{(\cdot)}^{xz}$ denotes a horizontal average. Here, however, we loosely refer to a streak simply as $\bar{u}$.  Noting that the essential attribute of a streak is the \emph{inflectional velocity anomaly} induced at its flanks that supports a streamwise-varying instability, this less restrictive definition seems more appropriate for flows with UMZs and VFs, for which the emergent $\overline{u}^{xz}$ profile itself may be inflectional and inviscidly unstable (cf. figure~\ref{fig:HOMOGENIZATION}b).} velocity field $\bar{u}$ is, itself, $\mathit{O}(1)$.  Similar arguments can be used to show that the effective Reynolds number arising in the momentum equation for the rolls ($\bar{v}$,$\bar{w}$) also is order unity. 

Two immediate deductions follow from this reasoning. Specifically, unless the rolls have an amplitude (e.g. a characteristic speed or circulation) that is much greater than $\mathit{O}(1/Re)$ as $Re\to\infty$, then: (i)~volume-mean viscous forces will be significant within the inertial domain, but this conclusion conflicts with the mean momentum balance upon volume averaging; and (ii)~the streak flow will be smoothly-varying, implying that the VF thickness (to the extent that a feature akin to a VF exists for such ECS) will \emph{not} scale with Reynolds number.  Thus, $\Delta_f/h$ will not appropriately decrease as the Reynolds number is increased, contradicting the empirically observed scaling behavior of VFs.

\begin{figure}
	\begin{center}
	\includegraphics[width=1.0\linewidth]{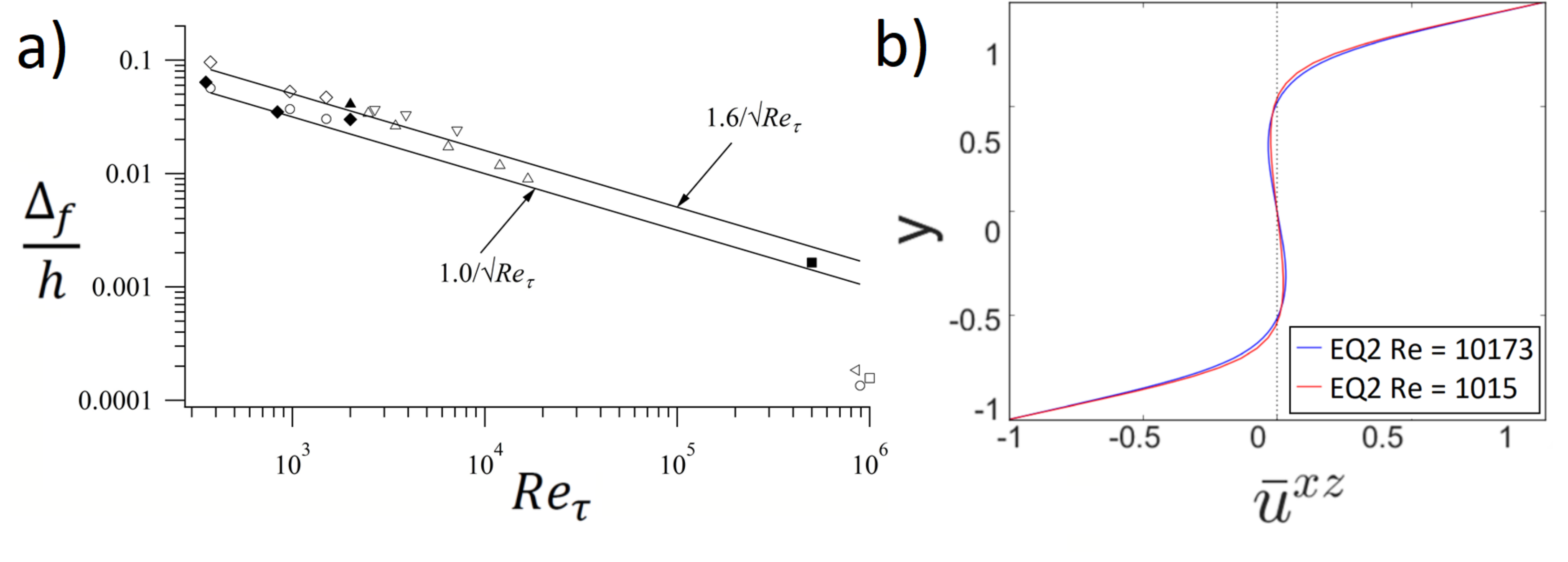}
	\caption{(a)~Ratio of the dimensional fissure width $\Delta_f$ to the BL height or channel half-height $h$ as a function of $Re_\tau$.  There is a clear power-law decrease in this ratio as $Re_\tau$ increases, with an exponent approximately equal to negative one-half.  Adapted from \cite{KlewickiIUTAM2013}.  (b)~The horizontally-averaged streamwise ($x$) velocity $\overline{u}^{xz}(y)$ associated with equilibrium solution EQ2, an \emph{upper-branch} state in plane Couette flow, at two different values of the Reynolds number $Re$.  Note that the profiles are nearly indistinguishable.  (Image courtesy of J. Gibson.)}
	\label{fig:FissureThickness}
	\end{center}
\end{figure}

To emphasize the second point, we reproduce in figure~\ref{fig:FissureThickness}(a) a plot from \cite{KlewickiIUTAM2013} showing a collection of experimentally obtained $\Delta_f/h$ estimates over a range of $Re_\tau$ spanning more than three decades; clearly, this ratio tends to zero as a power law of $Re_\tau$, with an approximate exponent of negative one-half.  In figure~\ref{fig:FissureThickness}(b), we show the horizontally-averaged (i.e. streamwise- and spanwise-mean) streamwise velocity $\overline{u}^{xz}(y)$ associated with EQ2, an \emph{upper-branch} equilibrium ECS in plane Couette flow, for two values of the Reynolds number that differ by an order of magnitude.  Contrary to a commonly held expectation, the strongly sheared (superficially ``fissure-like") regions near the upper and lower walls at $y=\pm 1$ do not continue to thin as the Reynolds number is increased.  Even for this upper-branch state, the $\overline{u}^{xz}(y)$ profile becomes independent of $Re$ provided this parameter is sufficiently large.  This scaling invariance is in complete accord with the assertion made by \cite{DeguchiHallJFM2014b} that all lower- and upper-branch ECS having $\mathit{O}(1)$ streamwise and spanwise wavenumbers (e.g. relative to the inverse domain size) will converge to VWI states as $Re\rightarrow\infty$. 

In the following section, we perform a large-$Re$ asymptotic analysis of the NS equations with the aim of deriving an SSP theory for \emph{inertial} ECS that can sustain UMZs and VFs 
in turbulent wall flows.  To do this, we seek a SSP in which the rolls, while still weak relative to the $\mathit{O}(1)$ streak flow, have an amplitude that is asymptotically larger than $\mathit{O}(1/Re)$.  Because the characteristic thickness of the shear layers induced by these roll motions is small relative to the roll diameter (or, equivalently, to the wall-normal spacing between adjacent VFs), the $x$-varying instability mode supported by the inflectional streak velocity profile has a wavelength that is commensurately small.  Thus, {\cb{unlike VWI theory}}, the proposed SSP is inherently multiscale in that the instability mode and roll/streak flow exhibit disparate spatial scales.  This scale separation is consistent with the observed flow physics of the inertial domain and, mathematically, implies that the ECS we construct cannot be obtained via solution of the VWI equations. 


\section{Asymptotic analysis of unbounded plane Couette flow}\label{sec:ASYMPTOTICS}
{\cb{In this section,}} we investigate whether counter-rotating streamwise rolls stacked in the wall-normal direction can differentially homogenize an imposed background shear flow.  More specifically, we imagine stacking in unbounded plane Couette flow infinitely many copies of the roll pattern depicted in figure~\ref{fig:Proposed}.  {\cb{The unbounded condition facilitates the analysis, as the spanwise roll (and instability-mode) velocity component does not vanish along the center of each fissure as it necessarily would along a no-slip channel wall. More importantly, we seek a self-sustaining mechanism that does not \emph{directly} invoke the presence of a wall, since UMZs and VFs are predominantly observed outboard of the near-wall Reynolds stress peak. Thus, solid boundaries are not included in our analysis except implicitly as a (remote) cause for the background shear (also see \S~\ref{sec:CONCLUSION}).}}

We choose to scale velocities by the friction velocity $u_\tau$, since the jump in streamwise flow speed across each fissure is a few times $u_\tau$ (figure~\ref{fig:conditional_u_omega}a,b), and lengths by $l_y$, the dimensional distance between adjacent -- and, in this construction, equispaced -- VFs.  The governing incompressible Navier--Stokes equations then can be expressed in dimensionless form as
\begin{eqnarray}\label{eq:NS}
	\partial_t \mathbf{u}+\mathbf{u}\cdot \nabla\mathbf{u}=-\nabla p +\frac{1}{Re}\nabla^2\mathbf{u},
\end{eqnarray}
where $\mathbf{u}=(u,v,w)$ and $p$ are the velocity vector and pressure, respectively, and incompressibility requires $\nabla\cdot\mathbf{u}=0$.  $Re\equiv u_\tau l_y/\nu$ is a Reynolds number defined using $l_y$ rather than the outer length scale $h$; that is, $Re\ne  Re_\tau$, a point we return to in \S~\ref{sec:CONCLUSION} (also see appendix~\ref{app:VFscaling}).  Given this non-dimensionalisation, the imposed background plane Couette flow is $u=y$, where $-\infty<y<\infty$.  Recently, \cite{HallJFM2018} constructed asymptotic ECS comprising an infinite wall-normal array of VWI states, each with a planar critical layer, firstly in unbounded Couette flow and subsequently in background shear flows with logarithmic profiles.  The theory we develop shares certain commonalities with the former construction but, as discussed further in \S~\ref{sec:CONCLUSION}, the distinctions highlighted in \S~\ref{sec:LIMITATION} remain apt.

\begin{figure}
	\begin{center}
		\includegraphics[width=0.8\linewidth]{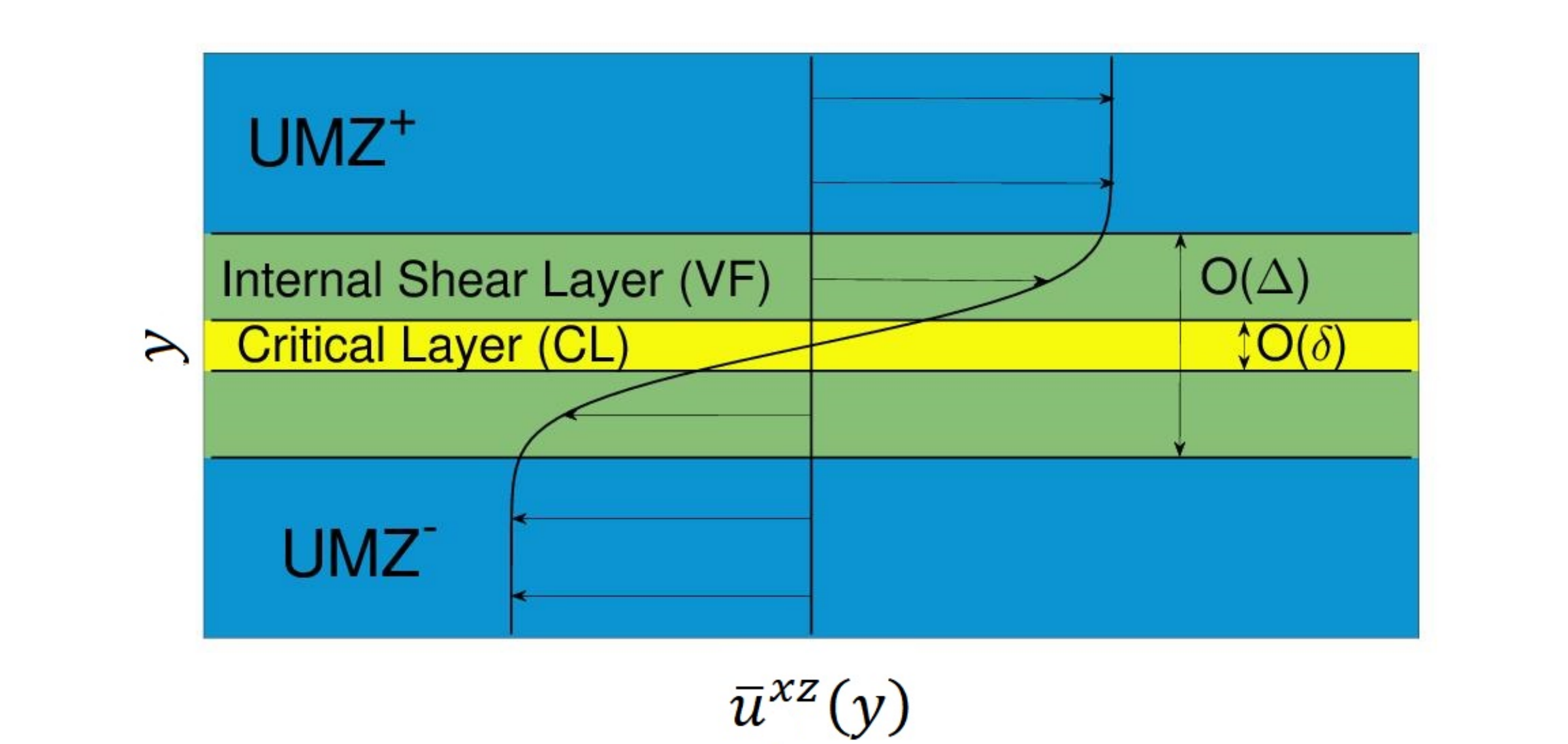}
		\caption{Schematic diagram of the hypothesized three-region asymptotic structure in the fissure-normal direction centred on a fissure located at $y=0$ {\cb{(cf. highlighted region in figure~\ref{fig:mean_force_balance}(b))}}. \textcolor{white}{Test test, why does this need to be on another line?}}
		\label{fig:3REGION}
	\end{center}	
\end{figure}

As in our earlier work \citep{BeaumePRE2015,ChiniPTRSA2017}, we decompose all flow fields into an $x$-mean plus a fluctuation about that mean to separate the streamwise-averaged roll and streak flow from the streamwise-varying instability mode.  Because the wavelength of the instability mode is small relative to the roll diameter and because the mode is refracted, a short scale not only in $x$ but also in $z$ is induced.  This rapid spanwise variation of the fluctuation is captured using a Wentzel--Kramers--Brillouin--Jeffreys (WKBJ) formalism introduced in \S~\ref{VF}.  Accordingly, we replace the streamwise average with an averaging operation that removes all variation in $x$ and fast variability both in time and in the spanwise direction.  Note, however, that we continue to denote mean fields with overbars and fluctuation fields with primes.  

{\cb{In the limit}} $Re\to\infty$ with $\mathit{O}(1/Re)\ll (\bar{v},\bar{w})\ll \mathit{O}(1)$ and $\bar{u}=\mathit{O}(1)$,
a three-region asymptotic structure emerges in the $y$ direction.  As shown in figure~\ref{fig:3REGION}, the bulk of the spatial domain is occupied by the UMZs.  Singularities arising in the largely dynamically-inviscid roll/streak flow within the UMZs are regularized by viscous forces acting within \emph{emergent} internal shear layers (VFs) of {\cb{dimensionless thickness $\Delta$}}, where $\Delta(Re)\to 0$ as $Re\to\infty$.  Within each fissure, the inflectional shear supports an inviscid (i.e. Rayleigh) instability of short streamwise and spanwise wavelength.  
Specifically, the streamwise wavenumber $\alpha=\breve{\alpha}/\Delta$, with $\breve{\alpha}=\mathit{O}(1)$ as $\Delta\to 0$.
{\cb{Consequently, $\alpha\to\infty$ as $Re\to\infty$, and}}
the Rayleigh mode is confined to the VF: the fluctuation fields decay exponentially with distance from the fissures, becoming transcendentally small (at asymptotically large $Re$) within the UMZs.  The Rayleigh mode, itself, exhibits a singularity that is viscously regularized within an even thinner critical layer (CL)  having dimensionless thickness $\mathit{O}(\delta)$, with $\delta/\Delta\to 0$ as $Re\to\infty$; see figure~\ref{fig:3REGION}. 

In the following subsections, we first analyze the flow within the UMZs adjacent to the VF centred on $y=0$ and subsequently the VF and its embedded CL.  Table~\ref{tab:scales} summarizes the scalings of the leading-order fields and highlights the dominant force balances arising in each of the three sub-regions for both the mean and fluctuating flow components.  {\cb{Collectively, these distinct dominant balances, along with the requirement that the mean roll vorticity at the edges of each CL smoothly match with that at the center of each collocated VF, ultimately determine the $Re$ dependencies of the VF and CL widths and of the roll and fluctuation amplitudes, as demonstrated below.}}  


\begin{table}
\begin{center}
	\begin{tabular}{ccccccc}\hline
	\multicolumn{2}{c}{\textbf{Domain}}& \multicolumn{2}{c}{\textbf{Dominant Contributions}}&\multicolumn{3}{c}{\textbf{ECS Component}}\\
			\hline
			\textbf{Region}&\textbf{Size}  & \textbf{Mean Terms} & \textbf{Fluctuations}&\textbf{S}&\textbf{R}&\textbf{F}\\ 
			\hline
			\textbf{UMZ}& $\mathit{O}(1)$  & NI, PG & E.S.T. & $\mathit{O}(1)$ & $\mathit{O}(\Delta^2)$ & E.S.T.\\
		 	\hline
			\textbf{VF}& $\mathit{O}(\Delta)$ & NI, VNF, PG & LI, PG & $\mathit{O}(1)$ & $\mathit{O}(\Delta^2)$ & $\mathit{O}(\Delta^3)$\\
			\hline
			\textbf{CL}& $\mathit{O}(\delta)$ & VNF, RSD  & LI, VNF,  PG & $\mathit{O}(\Delta)$ & $\mathit{O}(\Delta^2)$ & $\mathit{O}(\Delta^2)$\\
			\hline
	\end{tabular}
\caption{Summary of the scalings of the mean and fluctuation fields and the dominant terms arising in the relevant force balances in each of the three subdomains:  UMZ = uniform momentum zone; VF = vortical fissure (internal shear layer); and CL = critical layer.  Flow component:  S = streak; R = roll; and F = fluctuation.  Force:  NI = nonlinear inertia; LI = linearized inertia; PG = pressure gradient; VNF = viscous normal force; and RSD = Reynolds stress divergence.  As demonstrated by the analysis performed in \S~\ref{sec:ASYMPTOTICS}, the dimensionless VF thickness $\Delta=Re^{-1/4}$, while the CL thickness $\delta=Re^{-1/2}$.  E.S.T. = exponentially small terms.}
\label{tab:scales}
\end{center}
\end{table}	

\subsection{Uniform momentum zones}\label{UMZ}
The dynamics within the UMZs is governed by the two-dimensional (2D) but three-component (i.e. $x$-independent) NS equations, since the streamwise-varying fluctuation fields are exponentially small.  Thus, the momentum equations reduce to
\begin{eqnarray}\label{eq:UMZmean1}
\partial_t  \bar{u}+\left( \bar{\mathbf{v}}_{\bot}\cdot\nabla_\bot \right)\bar{u}&=&\frac{1}{Re}\nabla_\bot^2\bar{u},\label{eq:UMZ_umean}\\
\partial_t \bar{\mathbf{v}}_{\bot} + \left(\bar{\mathbf{v}}_{\bot}\cdot\nabla_\bot\right)\bar{\mathbf{v}}_{\bot}&=&-\nabla_\bot \bar{p}\,+\,\frac{1}{Re}\nabla_\bot^2\bar{\mathbf{v}}_{\bot},\label{eq:UMZ_vwmean}
\end{eqnarray}
where the $\bot$ subscript refers to the $y$--$z$ plane and the perpendicular (i.e. roll) velocity vector $\bar{\mathbf{v}}_\bot=(\bar{v},\bar{w})$.  From (\ref{eq:UMZ_umean}), it is clear that the mean streamwise velocity acts as a passive scalar within the UMZs, being advected by the rolls and diffused.  The rolls are not directly forced within the UMZs and therefore would decay in the absence of the forcing localized within the bounding VFs.  An immediate and significant physical implication is that the internal layers (sometimes referred to as ``interfaces" in the literature, e.g. see \cite{deSilvaJFM2017}) are not dynamically passive; rather, the driving agency for the staircase-like profiles of streamwise velocity is confined within the regions of concentrated vorticity. 

As noted in \S~\ref{sec:LIMITATION}, we insist that the \emph{dynamical} influence of viscosity on the streak and roll flow is weak, at least in a volume-averaged sense.  In particular, if the amplitude of the roll flow is denoted by $\bar{a}$, where $\bar{a}(Re)\to 0$ as $Re\to\infty$ (implying that the rolls are weak compared to the $\mathit{O}(1)$ streamwise streak flow), then we require the \emph{effective} Reynolds number $\bar{a}Re$ governing the dynamics of the rolls and streaks within the UMZs to become unbounded in the large-$Re$ limit.  It is well-known that in the presence of a steady cellular (2D) velocity field, a passive scalar will be homogenized in the limit of large Peclet number \citep{RhinesJFM1983}.   Since $\bar{a}Re\to\infty$, and given that we seek stacked ECS with steady rolls, the passive streak velocity field $\bar{u}$ therefore is differentially homogenized within regions of closed streamlines of the roll flow. 

\begin{figure}
	\begin{center}
		\includegraphics[width=1.0\linewidth]{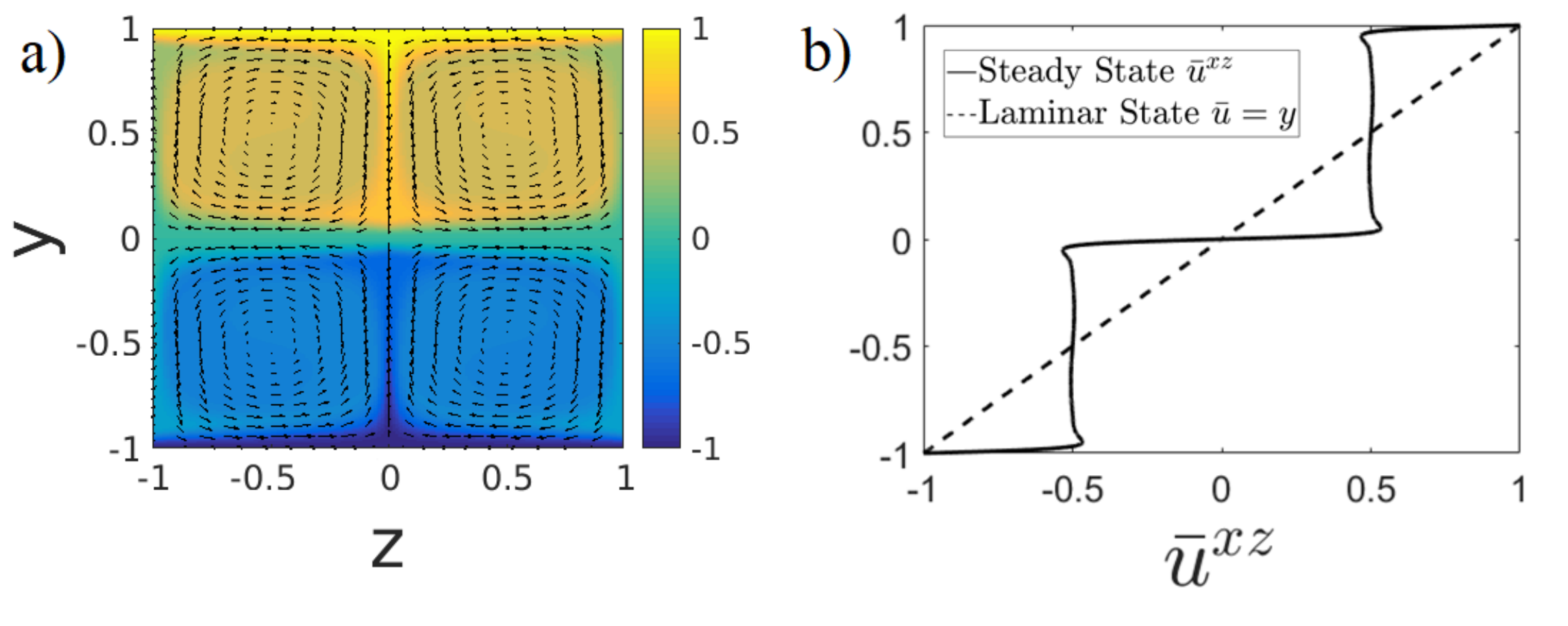}
		\caption{Differential homogenization of a background Couette flow (a,b) by {\cb{stacked, counter-rotating}} steady rolls (vector plot in (a), obtained from (\ref{eq:Vumz})--(\ref{eq:Wumz})) leading to the emergence of UMZs and an internal shear layer (b); i.e. an embedded VF.  This process is realized only for sufficiently large values of the \emph{effective} Reynolds number $\bar{a}Re$, where the roll amplitude $\bar{a}(Re)\to 0$ as $Re\to\infty$.  For example, in the Fourier--Chebyshev pseudospectral computations used to generate these results, $\bar{a}Re\approx 10^4$ (and $\bar{\Omega}_c\approx 7.35$.)}
		\label{fig:HOMOGENIZATION}
	\end{center}
\end{figure}

This process is clearly depicted in figure~\ref{fig:HOMOGENIZATION}.  The right-hand plot shows the initial and steady-state spanwise-averaged streamwise velocity profiles $\overline{u}^{xz}(y)$.  These profiles are obtained by numerically integrating (\ref{eq:UMZ_umean}) using a Fourier--Chebyshev pseudospectral algorithm with the prescribed steady roll velocity field given below in (\ref{eq:Vumz})--(\ref{eq:Wumz}) and plotted in figure~\ref{fig:HOMOGENIZATION}(a) for an effective Reynolds number $\bar{a}Re\approx 10^4$.  Of particular note is the emergence of an internal shear layer centred on the plane $y=0$.  

\cite{RhinesJFM1983} demonstrate that if the evolution is laminar (e.g. driven by a strictly steady 2D cellular velocity field acting on a scalar field exhibiting a uniform gradient at some initial time, say $t=0$), scalar homogenization in the large Peclet number ($Pe$) limit generically occurs in two stages.  During a time $t=\mathit{O}(Pe^{1/3})$, shear-augmented dispersion acts to homogenize the scalar field along streamlines.  The subsequent homogenization of the scalar field across streamlines occurs over a much longer time $t=\mathit{O}(Pe)$.  Crucially, diffusion is a leading-order process \emph{only} during the first phase:  in the present context, advection dominates perturbations due to diffusion,``locking" contours of constant $\bar{u}$ to the streamline pattern induced by the rolls, throughout the second phase as well as in the final steady state.  Plausibly, the time scale for homogenization may be reduced by turbulent mixing processes,  
but we make no judgment regarding the time taken for a turbulent trajectory to visit the neighborhood of the steady ECS we construct.  Nevertheless, we emphasize that within the UMZs viscous diffusion is a weak dynamical process relative to roll-induced advection for these ECS, in accord with the mean momentum balance outboard of the Reynolds stress peak (figure~\ref{fig:mean_force_balance}a).

For steady rolls and streaks within the homogenized core of the UMZs, the following asymptotic expansions are posited:
\begin{eqnarray}\label{eq:UMZscale}
		&u(x,y,z,t;Re)\sim \overline{u}_0(y,z)+\mbox{E.S.T.},\\
		&v(x,y,z,t;Re)\sim \bar{a}\overline{v}_2(y,z)+\ldots,\\
		&w(x,y,z,t;Re)\sim \bar{a}\overline{w}_2(y,z)+\ldots,\\
		&p(x,y,z,t;Re)\sim \bar{a}^2\overline{p}_4(y,z)+\ldots
\end{eqnarray}
The numeric subscripts refer to \emph{a posteriori} determined powers of the small parameter $\Delta$ defining $\bar{a}$, and E.S.T. denotes terms that are exponentially small in $Re$.  These terms include the $x$-varying fluctuation fields within the UMZs and transcendentally small corrections to the homogenized fields.  Indeed, not only is $\bar{u}$ homogenized but, in accord with the Prandtl--Batchelor theorem \citep{BatchelorJFM1956}, the steady $x$-mean $x$-vorticity $\bar{\Omega}$ also is uniform within regions foliated by steady closed streamlines on which viscous diffusion is weak relative to advection.  That is,
\begin{eqnarray}
\bar{\Omega}&\equiv&\partial_y\bar{w}-\partial_z\bar{v}\,\sim\,\nabla_\bot^2\left(\bar{a}\overline{\psi}_2\right)\,\sim\,\bar{a}\bar{\Omega}_c + \mbox{E.S.T.},\label{eq:POISSON}
\end{eqnarray}
where the (rescaled) roll streamfunction $\overline{\psi}_2(y,z)$ is defined by ($\overline{v}_2$,$\overline{w}_2$)=($-\partial_z\overline{\psi}_2$,$\partial_y\overline{\psi}_2$).  Unlike the constant core value of the $x$-mean streamwise velocity, which by symmetry must satisfy $\overline{u}_0\sim 1/2$ for $0<y<1$ and $\overline{u}_0=-1/2$ for $-1<y<0$, the (rescaled) homogenized value of the $x$-mean $x$-vorticity $\bar{\Omega}_c$, which fixes the precise value of the roll-induced circulation, is a primary unknown to be determined as part of the asymptotic analysis: effectively, $\bar{\Omega}_c$ is a nonlinear eigenvalue that will be shown to couple information from the various subdomains of the flow.  Nonetheless, as demonstrated in \cite{ChiniJFM2008}, (\ref{eq:POISSON}) is readily solved analytically on a rectangular domain of (asymptotically) known dimensions one unit in $y$ and $L_z/2$ units in $z$ (where $L_z$ is the prescribed spanwise periodicity length of the rolls), subject to $\overline{\psi}_2(y,z)\to 0$ as the rectangular cell boundaries are approached.  Using the given notation,
\begin{eqnarray}
\overline{v}_{2} &=& \bar{\Omega}_c\displaystyle \sum_{n=1, odd}^{\infty}\frac{2L_z}{(n\pi)^2}\left[\frac{\cosh[(2n\pi/L_z)(1/2-y)]}{\cosh(n\pi/L_z)}-1\right]\cos{\left(\frac{2n\pi z}{L_z}\right)},\label{eq:Vumz}\\
\overline{w}_{2} &=&\bar{\Omega}_c\displaystyle \sum_{n=1, odd}^{\infty}\frac{2L_z}{(n\pi)^2}\left[\frac{\sinh[(2n\pi/L_z)(1/2-y)]}{\cosh(n\pi/L_z)}\right]
\sin{\left(\frac{2n\pi z}{L_z}\right)}\label{eq:Wumz}.
\end{eqnarray}

To avoid the need for iteration in the subsequent solution algorithm, and noting that $\bar{\Omega}_c = O(1)$, it proves useful here to introduce the slightly modified small parameter 
$\tilde{\Delta} = \bar{\Omega}_c^{-1/4}\Delta$ 
and rescaled mean fields $\bar{v} = \bar{\Omega}_c\tilde{v}$, $\bar{w} = \bar{\Omega}_c\tilde{w}$ and $\bar{p} = \bar{\Omega}_c^2\tilde{p}$, where \emph{all} fields are (re-)expanded in asymptotic series in powers of $\tilde{\Delta}$ rather than $\Delta$. With this rescaling the governing equations in the UMZ become
\begin{eqnarray}\label{eq:UMZ_Umean}
\left( \tilde{\mathbf{v}}_{2\bot}\cdot\nabla_\bot \right)\bar{u}_{0}&=&
\tilde{\Delta}^2\nabla_\bot^2\overline{u}_{0},\label{eq:UMZ_UmeanU}\\
\left(\tilde{\mathbf{v}}_{2\bot}\cdot\nabla_\bot\right)\tilde{\mathbf{v}}_{2\bot}&=&-\nabla_\bot \tilde{p}_{4}\,+\,\tilde{\Delta}^2\nabla_\bot^2\tilde{\mathbf{v}}_{2\bot}.\label{eq:UMZ_UmeanVW}
\end{eqnarray}
Thus, with the understanding that $\bar{\Omega}_c$ is a to-be-determined constant, the steady roll velocity field within the UMZs is known, a key simplification.

%

\subsection{Vortical fissures}\label{VF}
In the asymptotic limit $Re\to\infty$, the differential homogenization {\cb{of $\bar{\Omega}$}} leads to jump discontinuities in {\cb{this}} field across the separatrices between adjacent roll cells. {\cb{Moreover, in addition to the jumps in $\bar{u}$ induced between stacked cells, streamwise velocity anomalies in the form of narrow jets are driven between neighboring roll pairs at each fixed $y$ location away from the fissures. These various}} discontinuities are smoothed by viscous forces and torques that act on the mean fields within asymptotically thin regions along the periphery of each cell.  Here, we focus on the emergent shear layer centred on $y=0$, but analogous considerations apply to all other VFs (located at $y=n$, for integer $n=\pm 1, \pm 2\ldots$).  Similar scalings also apply to the narrow jets centred on $z=m L_z/2$, for integer $m=0, \pm 1, \pm 2\ldots$, albeit with the roles of $y$ and $z$ interchanged and with the important distinction that, unlike the VFs, the jets are dynamically passive since in the present theory the $x$-varying fluctuations are exponentially small there.

The thickness of each VF follows from the usual laminar-BL scaling in which normal diffusion is balanced with advection, \emph{viz.} $\tilde{\Delta}=(\bar{a}\bar{\Omega}_c Re)^{-1/2}$.  For the VF centred on $y=0$, we therefore introduce a rescaled $y$ coordinate $\mathcal{Y}\equiv y/\tilde{\Delta}$, and we decompose all field variables into mean plus slowly-modulated fluctuation components:
\begin{eqnarray}\label{eq:VFexpansions}
\begin{bmatrix}
u(x,y,z,t;Re)\\
v(x,y,z,t;Re)\\
w(x,y,z,t;Re)\\
p(x,y,z,t;Re)
\end{bmatrix}
&\sim&
\begin{bmatrix}
\overline{\mathcal{U}}_0(\mathcal{Y},z)\\
\bar{a}\tilde{\Delta}\overline{\mathcal{V}}_3(\mathcal{Y},z)\\
\bar{a}\,\overline{\mathcal{W}}_2(\mathcal{Y},z)\\
\bar{a}^2\overline{\mathcal{P}}_4(\mathcal{Y},z)
\end{bmatrix}
+a'\mathcal{A}
\begin{bmatrix}
\hat{\mathcal{U}}_3(\mathcal{Y};z)\\
\hat{\mathcal{V}}_3(\mathcal{Y};z)\\
\hat{\mathcal{W}}_3(\mathcal{Y};z)\\
\hat{\mathcal{P}}_3(\mathcal{Y};z)
\end{bmatrix}
A(z)\mbox{e}^{i\left[\alpha (x-ct)+\theta(z/\tilde{\Delta})\right]}+c.c.,\qquad{}
\end{eqnarray}
where $c.c.$ denotes complex conjugate.
In these expansions, the fluctuations, which have a to-be-determined asymptotic magnitude $a'(Re)$ and $\mathit{O}(1)$ $z$-varying amplitude $\mathcal{A}A(z)$, are represented using a WKBJ approximation in which the fast phase $\theta(z/\tilde{\Delta})\equiv\Theta(z)/\tilde{\Delta}$ and the rescaled spanwise wavenumber $\tilde{\beta}\equiv \partial_z\Theta=\mathit{O}(1)$.  We also define the $O(1)$ streamwise wavenumber $\tilde{\alpha} = \alpha\tilde{\Delta}$ and, for subsequent reference, note that 
$\breve{\alpha} = \tilde{\alpha}\bar{\Omega}_c^{1/4}$, 
i.e. the $O(1)$ streamwise wavenumber scaled by $\Delta$ rather than by $\tilde{\Delta}$. 
The $\mathit{O}(1)$ phase speed $c$ is strictly real for neutral Rayleigh modes implicated in a steady SSP and vanishes only for the VF at $y=0$.
Although the real scalar $\mathcal{A}$ could be absorbed into the definition of the amplitude function $A(z)$, it proves convenient to explicitly retain this factor as a control parameter.  (That is, we take $\mathcal{A}$ and the rescaled, $\mathit{O}(1)$ streamwise wavenumber $\breve{\alpha}$ as control parameters for the Rayleigh mode and self-consistently determine the slowly-varying spanwise wavenumber $\tilde{\beta}(z)$ and amplitude function $A(z)$.)  To disentangle $\mathcal{A}$ from $A(z)$, we normalize the latter such that $A(0)=1$.  An additional normalization condition will be specified below to distinguish the $\mathcal{Y}$- and $z$-varying eigenfunction from the amplitude, thereby rendering the decomposition of the fluctuation fields in (\ref{eq:VFexpansions}) unique.


The scaling of the mean streamwise and spanwise velocity components in (\ref{eq:VFexpansions}) ensures smooth matching with the flow in the adjacent UMZs is possible, while the scaling of the mean fissure-normal velocity follows from incompressibility.   In the proposed configuration, no physical (i.e. no-slip) boundary exists along the horizontal planes separating rows of stacked counter-rotating rolls.  Consequently, the leading-order mean spanwise velocity component within the VF is not sheared; i.e. $\partial_\mathcal{Y}\overline{\mathcal{W}}_2=0$, hence $\overline{\mathcal{W}}_2=\overline{\mathcal{W}}_2(z)$, only.  Through the mean incompressibility condition, this ansatz implies that within the fissure $\overline{\mathcal{V}}_3(\mathcal{Y},z)$ varies linearly with the normal coordinate $\mathcal{Y}$.  A second immediate consequence is that the roll vorticity will have the same asymptotic size, namely $\mathit{O}(\bar{a})$, within the VF and adjacent UMZs, ensuring smooth matching of this field is possible.  To close the global vorticity budget, however, a $\mathcal{Y}$-dependent correction $\bar{a}\tilde{\Delta}\overline{\mathcal{W}}_3(\mathcal{Y},z)$ must be appended to the expansion for the mean spanwise velocity component in (\ref{eq:VFexpansions}); see \cite{HarperJFM1963}.  Finally, in contrast to the mean fields, the fluctuations are isotropic within the fissure and thus each fluctuation field has the same asymptotic magnitude $a'$.  (We find that $a'\le\bar{a}$ as $Re\to\infty$ for sensible physical balances to be realized in the mean equations.)  Recalling that the streamwise wavelength of the fluctuation fields is commensurate with ${\Delta}$, the VF thickness, the fluctuations consequently decay exponentially away from the centre of the fissure.

\subsubsection{Viscous mean dynamics:  Childress cell problem}\label{sec:Childress}

Substituting (\ref{eq:VFexpansions}) into the incompressibility condition and the NS equations, applying the streamwise/fast-phase averaging operation and collecting terms at leading order in $\tilde{\Delta}$ yields
\begin{eqnarray}\label{eq:VFmeantilde}
\partial_{\mathcal{Y}}\tilde{\mathcal{V}}_3\,+\,\partial_z\tilde{\mathcal{W}_2}&=&0,\label{eq:VFmeanCONT}\\
\tilde{\mathcal{V}}_3\partial_{{\mathcal{Y}}}\overline{\mathcal{U}}_0+\tilde{\mathcal{W}}_2\partial_z\overline{\mathcal{U}}_0&=& \partial_{{\mathcal{Y}}}^2\overline{\mathcal{U}}_0,\label{eq:VFmeanU}\\
\partial_{\mathcal{Y}}\tilde{\mathcal{P}}_4&=&0,\label{eq:VFmeanP}\\
\tilde{\mathcal{W}}_2\partial_z\tilde{\mathcal{W}}_2&=&-\partial_z\tilde{\mathcal{P}}_4,\label{eq:VFmeanW}\\
\tilde{\mathcal{V}}_3\partial_{{\mathcal{Y}}}\left(\partial_{\mathcal{Y}}\tilde{\mathcal{W}}_3\right)+\tilde{\mathcal{W}}_2\partial_z\left(\partial_{\mathcal{Y}}\tilde{\mathcal{W}}_3\right)&=&-\,\frac{\partial_{\mathcal{Y}}^2\left(\overline{\mathcal{V}'_3\mathcal{W}'_3}\right)}{\bar{\Omega}_c^2}+\,\partial_{{\mathcal{Y}}}^2\left(\partial_{\mathcal{Y}}\tilde{\mathcal{W}}_3\right),\label{eq:VFmeanOMEGA}
\end{eqnarray}
where $(\overline{\mathcal{V}}_3,\overline{\mathcal{W}}_{2,3}) = \bar{\Omega}_c(\tilde{\mathcal{V}}_3,\tilde{\mathcal{W}}_{2,3})$ and $\overline{\mathcal{P}}_4=\bar{\Omega}_c^2\tilde{\mathcal{P}}_4$. 
The key simplification to the mean flow equations in this region is that asymptotic matching enables the leading-order roll flow $(\tilde{\mathcal{V}}_3,\tilde{\mathcal{W}}_2)$ to be obtained by extrapolating the UMZ roll solution to the VF. Specifically, the tangential component of the roll flow in the fissure is obtained simply by evaluating (\ref{eq:Wumz}) at $y = 0$. Using the incompressiblity constraint (\ref{eq:VFmeantilde}) and the symmetry condition $\tilde{\mathcal{V}}_3(\mathcal{Y} = 0,z) = 0$ to determine a constant of integration, the normal flow component $\tilde{\mathcal{V}}_3$ also can be determined, yielding
\begin{eqnarray}\label{eq:VWtildeVF}
&\tilde{\mathcal{V}}_{3}({\mathcal{Y}},z) = -{\mathcal{Y}} \displaystyle \sum_{n=1, odd}^{\infty}\left(\frac{4}{n\pi}\right)\tanh(\frac{n\pi}{L_z})\cos\left(\frac{2n\pi z}{L_z}\right),\label{eq:VtildeVF}\\
&\tilde{w}_{2}(y=0,z) = \tilde{\mathcal{W}}_2(z) = \displaystyle \sum_{n=1, odd}^{\infty}\left(\frac{2L_z}{(n\pi)^2}\right)\tanh(\frac{n\pi}{L_z})\sin\left(\frac{2n\pi z}{L_z}\right).\label{eq:WtildeVF}
\end{eqnarray}
The leading-order roll flow within the VF (\ref{eq:VtildeVF})--(\ref{eq:WtildeVF}) satisfies the mean equations (\ref{eq:VFmeanCONT}), (\ref{eq:VFmeanP}) and (\ref{eq:VFmeanW}), with $\tilde{\mathcal{P}}_4(z) = \tilde{p}_4(y$=$0,z)$ for matching with the adjacent UMZs.

Given (\ref{eq:VtildeVF}) and (\ref{eq:WtildeVF}), the ($\perp$) velocity field that advects the leading-order streak flow $(\overline{\mathcal{U}}_0)$ and $x$-vorticity $(\partial_\mathcal{Y}\tilde{\mathcal{W}}_3)$ within the fissure is known, and the equations for these fields effectively linearise. Analogous considerations apply to the viscous jet regions around the remainder of the roll-cell periphery.  By choosing to scale $a' = \tilde{\Delta}\overline{a}$, the torque provided by the gradient of the fluctuation-induced Reynolds stress divergence $-\partial_\mathcal{Y}(\overline{\mathcal{V}_3'\mathcal{W}_3'})$ arises consistently in (\ref{eq:VFmeanOMEGA}) and seemingly has the potential to generate $x$-vorticity within the VF and, thence, to drive roll motions within the UMZ. Subsequently, however, it is confirmed that $\mathcal{V}_3'$ and $\mathcal{W}_3'$ are out of phase for a (3D) neutral Rayleigh mode. Consequently, this correlation vanishes and no driving of the rolls by this mechanism is realized. Instead, as in VWI theory, the driving to sustain the roll motions arises from a critical-layer phenomenon (see \S~\ref{sec:CL} and \S~\ref{subsec:energybalance}).

The vorticity equation (\ref{eq:VFmeanOMEGA}) must be solved subject to the matching condition $\partial_\mathcal{Y}\tilde{\mathcal{W}}_3\rightarrow -1$ as $\mathcal{Y}\rightarrow\infty$ (e.g. for matching with the upper UMZ in $0\le z\le L_z/2$). The appropriate boundary condition on this vorticity component as $\mathcal{Y}\rightarrow 0^+$ is not obvious \emph{a priori}, but will be determined in \S~\ref{sec:CL}. Indeed, in the absence of driving from the Reynolds stress term in (\ref{eq:VFmeanOMEGA}), this boundary condition will be shown to give rise to the forcing that sustains the rolls, revealing that the CL plays a key role in maintaining the staircase-like profile of streamwise velocity in the proposed asymptotic SSP.

\begin{figure}
\begin{center}
\includegraphics[width=1.0\linewidth]{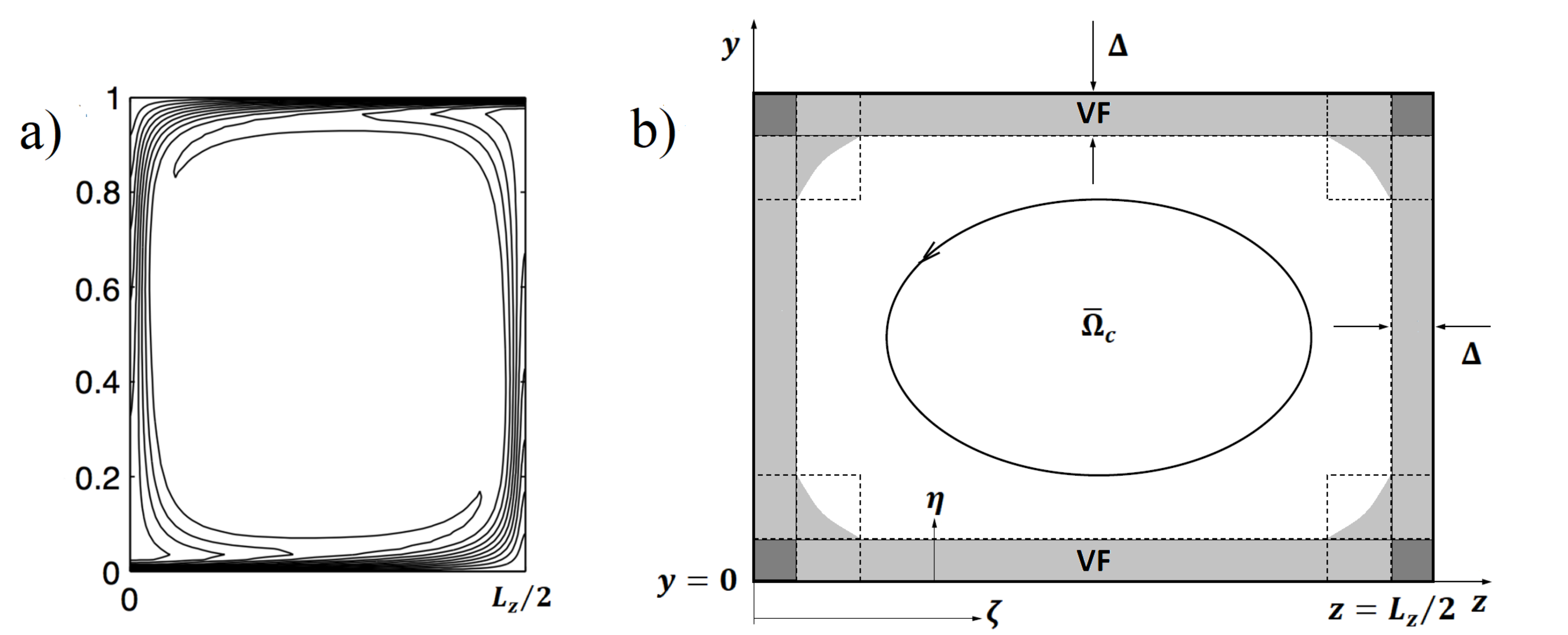}
\caption{Formulation of Childress cell problem, as adapted from \cite{ChiniPoF2009}. (a)~Contour plot of $\overline{u}_0(y,z)$. (b)~Corresponding multiregion asymptotic structure of a steady roll cell as $Re\to\infty$. The white region is the dynamically inviscid vortex core. The light grey regions indicate the $O(\Delta)$ thick VFs as well as $O(\Delta)$ thick positive and negative streamwise jets. The streamwise-averaged streamwise velocity is passively advected through the outer corner regions also indicated by the light grey shading.  (The viscous inner corner regions highlighted in dark grey also are dynamically passive.) $\zeta$ is a stretched arc-length coordinate running around the cell perimeter, and $\eta$ is a scaled coordinate measuring distance normal to the VFs and jets.}
\label{fig:CHILDRESS}
\end{center}
\end{figure}

The mean streamwise momentum equation (\ref{eq:VFmeanU}) must be solved subject to the symmetry conditions $\overline{\mathcal{U}}_0 \rightarrow 1/2$ as $\mathcal{Y}\rightarrow\infty$ (for the UMZ located in $0<y<1$) and $\overline{\mathcal{U}}_0 \rightarrow 0$ as $\mathcal{Y}\rightarrow 0$. Identical equations and boundary conditions apply within the VF centred on $y=1$ upon replacing $\tilde{\Delta}\mathcal{Y} \rightarrow 1-\tilde{\Delta}\mathcal{Y}$ and $z \rightarrow L_z/2-z$. The same steady advection--diffusion equation also applies within the streamwise jets centred on $z=0$ and $z = L_z/2$, with the roles of $y$ and $z$ (and $\tilde{\mathcal{V}}_3$ and $\tilde{\mathcal{W}}_2$) interchanged. The resulting problem for $\overline{\mathcal{U}}_0$ is formally identical to that for the temperature field in steady 2D Rayleigh-B\'enard convection (RBC) in the limit of asymptotically large Rayleigh number.  This observation provides an interesting and potentially useful connection between coherent structures arising in strongly nonlinear convection and wall-bounded shear flows. (Of course, in RBC, the cellular flow is driven by buoyancy torques acting within vertical plumes, while in the proposed SSP, the roll flow is driven by Reynolds stresses arising from the nonlinear interaction of a streamwise-varying shear instability mode confined to the horizontal fissures.) 
As demonstrated in \cite{ChiniPoF2009}, the advected and diffused scalar field can be determined by formulating a \emph{Childress cell problem} {\cb{\citep{ChildressALPHA1979}}}, in which the fissures and jets (or thermal boundary layers and plumes in the RBC context) around each cell are stitched together to form a connected domain{\cb{; see figure \ref{fig:CHILDRESS}. Here, we simply lift the key results from \cite{ChiniPoF2009}; see appendix~\ref{app:ChildressDetails} for further details.}}  In particular, solution of the Childress cell problem yields the shear $\tilde{\lambda}(z) \equiv \partial_{\mathcal{Y}} \overline{\mathcal{U}}_0(\mathcal{Y}=0,z)$ induced by the mean streamwise velocity component at the center of the fissure.

\begin{figure}
\begin{center}
\includegraphics[width=0.75\linewidth]{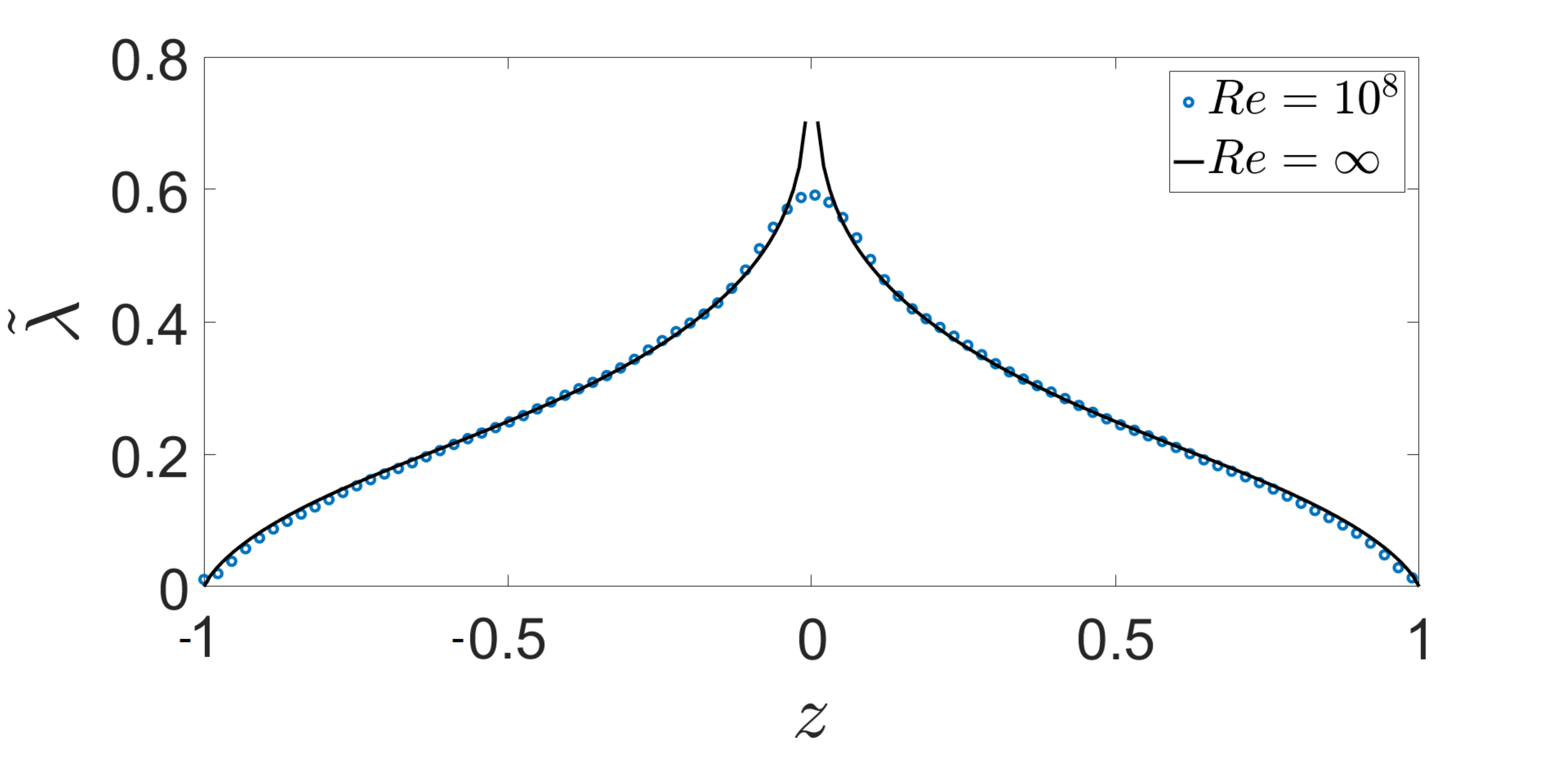}
\caption{Normalized streak-induced shear $\tilde{\lambda}(z)\equiv \partial_\mathcal{Y}\overline{\mathcal{U}}_0(0,z)$ at the mid-plane of the VF obtained from the solution of the Childress cell problem (i.e. in the limit $Re\rightarrow\infty$, solid curve) and from the numerical solution of the full advection--diffusion equation (\ref{eq:UMZ_Umean}) for finite $Re$ (dotted curve).} 
\label{fig:ShearCoreVorticity}
\end{center}
\end{figure}

The solution for $\tilde{\lambda}(z)$ obtained from the Childress cell problem, i.e. in the limit $Re$$\rightarrow$$\infty$, is compared in figure \ref{fig:ShearCoreVorticity} to that obtained by numerically solving the advection--diffusion equation (\ref{eq:UMZ_UmeanU}) for small but finite $\tilde{\Delta}^2$ (corresponding to large but finite $Re$). The two profiles agree closely except at $z=0$, where diffusion heals a passive singularity arising in the solution of the Childress cell problem. This validation enables the finite-$Re$ data (i.e. with viscous regularization) to be used in the calculation of the fluctuation velocities within the VF, which is considered next.  This strategy proves to be more convenient and robust than attempting to directly reconstruct $\overline{\mathcal{U}}_0(\mathcal{Y},z)$ from the solution of the Childress cell problem.

\subsubsection{Inviscid fluctuation dynamics:  Rayleigh, Eikonal and amplitude equations}
Recalling the WKBJ ansatz in (\ref{eq:VFexpansions}), the leading-order equations for the fluctuation fields within the VF are
\begin{eqnarray}\label{eq:VFfluc3d}
	i\tilde{\alpha}\hat{\mathcal{U}}_3\,+\,\partial_{\mathcal{Y}}\hat{\mathcal{V}}_3\,+\,i\tilde{\beta}\hat{\mathcal{W}}_3&=0,\label{eq:VFflucCONT}\\
	i\tilde{\alpha}\left(\overline{\mathcal{U}}_0-c\right)\hat{\mathcal{U}}_3\,+\,\hat{\mathcal{V}}_3\partial_{\mathcal{Y}}\overline{\mathcal{U}}_0\,+\,i\tilde{\alpha}\hat{\mathcal{P}}_3&=0,\label{eq:VFflucU}\\
	i\tilde{\alpha}\left(\overline{\mathcal{U}}_0-c\right)\hat{\mathcal{V}}_3\,+\,\partial_\mathcal{Y}\hat{\mathcal{P}}_3&=0,\label{eq:VFflucV}\\
	i\tilde{\alpha}\left(\overline{\mathcal{U}}_0-c\right)\hat{\mathcal{W}}_3\,+\,i\tilde{\beta}\hat{\mathcal{P}}_3&=0.\label{eq:VFfluc3dW}
\end{eqnarray}
From (\ref{eq:VFflucCONT})--(\ref{eq:VFfluc3dW}), it can be deduced that $\hat{\mathcal{V}}_3$ and $\hat{\mathcal{W}}_3$ are $\pi/2$ out of phase, confirming that the correlation $\overline{\mathcal{V}_3'\mathcal{W}_3'}=0$ in (\ref{eq:VFmeanOMEGA}).
Utilizing Squire's transformation, this 3D system can be reduced to a 2D system. Specifically, adding $\tilde{\alpha}\times$(\ref{eq:VFflucU}) to $\tilde{\beta}\times$(\ref{eq:VFfluc3dW}) and dividing the sum by $\tilde{\alpha}$ yields
\begin{eqnarray}\label{eq:VFfluc2d}
	i \tilde{k} \hat{\mathcal{U}}_{3r}\,+\,\partial_{\mathcal{Y}}\hat{\mathcal{V}}_3&=0,\label{eq:VFfluc2dCONT}\\
	i \tilde{k} \left(\overline{\mathcal{U}}_0-c\right)\hat{\mathcal{U}}_{3r}\,+\,\hat{\mathcal{V}}_3\partial_{\mathcal{Y}}\overline{\mathcal{U}}_0\,+\,i \tilde{k}\hat{\Pi}_3&=0,\label{eq:VFfluc2du}\\
	i \tilde{k} \left(\overline{\mathcal{U}}_0-c\right)\hat{\mathcal{V}}_3\,+\,\partial_\mathcal{Y}\hat{\Pi}_3&=0,\label{eq:VFfluc2dv}
\end{eqnarray}
along with a decoupled but important equation for $\hat{\mathcal{W}}_{3r}$,
\begin{equation}\label{eq:VFfluc2dw}
    \hat{\mathcal{W}}_{3r}= \frac{1}{i\tilde{k}}\frac{\tilde{\beta}}{\tilde{\alpha}}\frac{\hat{\mathcal{V}}_{3}\partial_\mathcal{Y}\overline{\mathcal{U}}_0}{(\overline{\mathcal{U}}_0-c)},
\end{equation}
where
\begin{eqnarray}
	\tilde{k}=\sqrt{\tilde{\alpha}^2+\tilde{\beta}^2},\qquad& \tilde{k}\hat{\mathcal{P}}_3=\tilde{\alpha}\hat{\Pi}_3, \label{eq:transformations}\\
	\tilde{k}\hat{\mathcal{U}}_{3r}=\tilde{\alpha}\hat{\mathcal{U}}_3+\tilde{\beta}\hat{\mathcal{W}}_3\qquad \Longleftrightarrow& \qquad \tilde{k}\hat{\mathcal{U}}_3=\tilde{\alpha}\hat{\mathcal{U}}_{3r}-\tilde{\beta}\hat{\mathcal{W}}_{3r},\label{eq:transformations2}\\
	\tilde{k}\hat{\mathcal{W}}_{3r}=\tilde{\alpha}\hat{\mathcal{W}}_3-\tilde{\beta}\hat{\mathcal{U}}_3\quad\;\; \Longleftrightarrow& \quad\quad \tilde{k}\hat{\mathcal{W}}_3=\tilde{\beta}\hat{\mathcal{U}}_{3r}+\tilde{\alpha}\hat{\mathcal{W}}_{3r}.\label{eq:transformations3}
\end{eqnarray}
Equations (\ref{eq:VFfluc2dCONT})--(\ref{eq:VFfluc2dv}) comprise a linear homogeneous eigensystem and are recognizable as the equations governing the stability of plane parallel shear flows to small-amplitude 2D disturbances in the inviscid limit; indeed, (\ref{eq:VFfluc2dCONT})--(\ref{eq:VFfluc2dv}) may be collapsed into Rayleigh's stability equation, albeit here with a base flow $\overline{\mathcal{U}}_0$ that depends parametrically on the spanwise coordinate $z$.  Since a neutral mode is sought and the associated wave phase speed is set by symmetry considerations (i.e. $c=n$ for the VF centred at $y=n$, where $n=0,\pm1, \pm2 \ldots$), the total horizontal wavenumber $\tilde{k}$ may be taken as the eigenvalue. Once $\tilde{k}$ is computed, the spanwise wavenumber $\tilde{\beta}$ may be determined. Note that this 1D eigenvalue problem must be solved over the domain $z \in [-L_z/2, L_z/2]$, since the streak field $\overline{\mathcal{U}}_0$ varies (slowly) with $z$. The shape of the resulting neutral mode is given by the eigenfuction. To determine the slowly $z$-varying amplitude, however, the fluctuation equations must be analyzed at higher order.

To that end, collecting terms at the next order in $\tilde{\Delta}$ yields
\begin{equation}
\begin{aligned}
B\left[i\tilde{\alpha}\hat{\mathcal{U}}_4\,+\,\partial_{\mathcal{Y}}\hat{\mathcal{V}}_4\,+\,i\tilde{\beta}\hat{\mathcal{W}}_4\right]&= - \partial_z (A\hat{\mathcal{W}_3}),\\
B\left[i\tilde{\alpha}\left(\overline{\mathcal{U}}_0-c\right)\hat{\mathcal{U}}_4\,+\,\hat{\mathcal{V}}_4\partial_{\mathcal{Y}}\overline{\mathcal{U}}_0\,+\,i\tilde{\alpha}\hat{\mathcal{P}}_4\right]&= -(i\tilde{\alpha})\overline{\mathcal{U}}_1A\hat{\mathcal{U}}_3-A\hat{\mathcal{V}}_3\partial_{\mathcal{Y}}\overline{\mathcal{U}}_1-A\hat{\mathcal{W}_3}\partial_z\overline{\mathcal{U}}_0,\\
B\left[i\tilde{\alpha}\left(\overline{\mathcal{U}}_0-c\right)\hat{\mathcal{V}}_4\,+\,\partial_{\mathcal{Y}}\hat{\mathcal{P}}_4\right]&=-(i\tilde{\alpha})\overline{\mathcal{U}}_1A\hat{\mathcal{V}}_3,\\
B\left[i\tilde{\alpha}\left(\overline{\mathcal{U}}_0-c\right)\hat{\mathcal{W}}_4\,+\,i\tilde{\beta}\hat{\mathcal{P}}_4\right]&=-(i\tilde{\alpha})\overline{\mathcal{U}}_1A\hat{\mathcal{W}}_3-\partial_z(A\hat{\mathcal{P}}_3),
\end{aligned}
\end{equation}
where $B(z)$ is the $z$-varying amplitude of the $O(\tilde{\Delta}^4)$ fluctuation fields. The Fredholm alternative solvability condition along with strategic use of the fluctuation incompressibility constraint to eliminate terms proportional to the unknown mean-field correction $\overline{\mathcal{U}}_1$ then yields an amplitude equation of the form
\begin{eqnarray}\label{eq:AmpEqn}
	a_1\frac{dA}{dz}\,+\,a_2 A&=&0,
\end{eqnarray}
where
\begin{equation}
\begin{aligned}
a_1 = &{\cb{\int_{-\infty}^{\infty}}}\left(\frac{i\hat{\mathcal{V}}^*_3}{\overline{\mathcal{U}}_0-c}\right)\left[\frac{\tilde{\beta}}{\tilde{k}}\partial_{\mathcal{Y}}\hat{\Pi}_3-\partial_{\mathcal{Y}}\overline{\mathcal{U}}_0\left(\frac{\tilde{\beta}}{\tilde{k}}\hat{\mathcal{U}}_{3r}+\frac{\tilde{\alpha}}{\tilde{k}}\hat{\mathcal{W}}_{3r}\right)-\left(\overline{\mathcal{U}}_0-c\right)\partial_{\mathcal{Y}}\left(\frac{\tilde{\beta}}{\tilde{k}}\hat{\mathcal{U}}_{3r}+\frac{\tilde{\alpha}}{\tilde{k}}\hat{\mathcal{W}}_{3r}\right)\right]d{\mathcal{Y}},\\
a_2 = &{\cb{\int_{-\infty}^{\infty}}}\left(\frac{i\hat{\mathcal{V}}^*_3}{\overline{\mathcal{U}}_0-c}\right)\Bigg[\tilde{\beta}\partial_z\left(\frac{\partial_{\mathcal{Y}}\hat{\Pi}_3}{\tilde{k}}\right)-\left(\partial_{\mathcal{Y}}\overline{\mathcal{U}}_0\right)\partial_z\left(\frac{\tilde{\beta}}{\tilde{k}}\hat{\mathcal{U}}_{3r}+\frac{\tilde{\alpha}}{\tilde{k}}\hat{\mathcal{W}}_{3r}\right)+\partial_z(\partial_{\mathcal{Y}}\overline{\mathcal{U}}_0)\left(\frac{\tilde{\beta}}{\tilde{k}}\hat{\mathcal{U}}_{3r}+\frac{\tilde{\alpha}}{\tilde{k}}\hat{\mathcal{W}}_{3r}\right)\\
&-\left(\overline{\mathcal{U}}_0-c\right)\partial_z\left(\partial_{\mathcal{Y}}\left(\frac{\tilde{\beta}}{\tilde{k}}\hat{\mathcal{U}}_{3r}+\frac{\tilde{\alpha}}{\tilde{k}}\hat{\mathcal{W}}_{3r}\right)\right)+\left(\partial_z\overline{\mathcal{U}}_0\right)\partial_{\mathcal{Y}}\left(\frac{\tilde{\beta}}{\tilde{k}}\hat{\mathcal{U}}_{3r}+\frac{\tilde{\alpha}}{\tilde{k}}\hat{\mathcal{W}}_{3r}\right)\Bigg]d\mathcal{Y}\\
\end{aligned}
\end{equation}
{\cb{and}} an asterisk denotes complex conjugation{\cb{. The apparent singularities arising from the prefactor $(\overline{\mathcal{U}}_0-c)^{-1}$ in the expressions for $a_1$ and $a_2$ can be removed by utilizing (\ref{eq:VFfluc2dCONT})--(\ref{eq:VFfluc2dv}), yielding}}
\begin{equation}\label{eq:a1234}
\begin{aligned}
a_1 & =2\tilde{\beta} \int_{-\infty}^{\infty}\hat{\mathcal{V}}_3\hat{\mathcal{V}}^*_3d\mathcal{Y},\\
a_2 & = \int_{-\infty}^{\infty}\Bigg\{2\tilde{\beta}(\partial_z\hat{\mathcal{V}}_3)\hat{\mathcal{V}}_3^*+(\partial_z\tilde{\beta})\hat{\mathcal{V}}_3\hat{\mathcal{V}}_3^* -2\frac{\tilde{\beta}}{\tilde{k}^2}\partial_{\mathcal{Y}}\left(\frac{\partial_z\overline{\mathcal{U}}_0}{\overline{\mathcal{U}}_0-c}\right)(\partial_{\mathcal{Y}}\hat{\mathcal{V}}_3)\hat{\mathcal{V}}_3^*\\
&\qquad\qquad+\,2\frac{\tilde{\beta}}{\tilde{k}^2}\partial_{\mathcal{Y}}\left(\frac{\partial_z\overline{\mathcal{U}}_0}{\overline{\mathcal{U}}_0-c}\right)\left[\frac{\hat{\mathcal{V}}_3\hat{\mathcal{V}}_3^*\partial_{\mathcal{Y}}\overline{\mathcal{U}_0}}{\overline{\mathcal{U}}_0-c}\right]\Bigg\}d\mathcal{Y}. 
\end{aligned}
\end{equation}
Upon noting $\overline{\mathcal{U}}_0\sim \tilde{\lambda}\mathcal{Y}$ as $\mathcal{Y}\rightarrow 0$, it can be confirmed that these expressions for the $z$-dependent coefficients are not singular.

The amplitude equation (\ref{eq:AmpEqn}) can be integrated to obtain
\begin{eqnarray}
&A(z) = \text{exp}\left(-\int_{0}^{z}\frac{a_2(\hat{z})}{a_1(\hat{z})}\,d\hat{z}\right), \label{eq:A}
\end{eqnarray}
where $A(z) = A(-z)$, i.e. $A$ is even about $z=0$, and the constant of integration has been subsumed into the constant $\mathcal{A}$ arising in the WKBJ ansatz. This solution breaks down at those $z$ locations where $a_1\rightarrow 0$. In particular, the coefficient 
$a_1\rightarrow 0^+$ while $a_2<0$ remains finite as $\tilde{\beta}\rightarrow 0^+$, as can be corroborated using (\ref{eq:a1234}). 
Thus, there is a curve (a line in the $x$--$z$ plane in the present context), termed a \emph{caustic}, on which the amplitude $A(z)$ formally becomes unbounded. The caustic separates the flow within the VF into a region with finite-amplitude fluctuations, termed the ``illuminated'' region, and a region in which the fluctuations are exponentially small, termed the ``shadow'' region. A more careful analysis (see appendix \ref{app:Caustic}) confirms that within an asymptotically small neighborhood of each caustic the amplitude is viscously regularized and thus does not become unbounded. Moreover, the spanwise fluctuation velocity is suppressed at the caustic since the spanwise fluctuation pressure gradient is weak there relative to its magnitude in the illuminated region, enabling the incident Rayleigh mode to two-dimensionalize prior to being reflected. Upon reflection, the rays leaving each caustic experience a $\pi/2$ phase shift in $z$.  Finally, it can be shown that the contribution to the roll-flow energy budget (performed in \S~\ref{subsec:energybalance}) from this neighborhood of each caustic is asymptotically small (again, because $w'$ is suppresed there) and therefore can be neglected from the integral (\ref{eq:OmegEqn}) used to calculate $\bar{\Omega}_c$.

\subsection{Critical layers}\label{sec:CL}
Although the rotated neutral-mode velocity components $\hat{\mathcal{U}}_{3r}$ and $\hat{\mathcal{V}}_{3}$ are regular 
{\cb{for all $\mathcal{Y}$,}} the rotated spanwise fluctuation velocity $\hat{\mathcal{W}}_{3r}$ is singular as $\mathcal{Y}\to 0$, as evident from inspection of (\ref{eq:VFfluc2dw}). Thus, a critical layer (CL) emerges within the VF, where the horizontal mean streamwise flow speed  matches the phase speed (i.e. 0, for the VF centred at $y = 0$) of the instability mode. Within this CL, the $\mathcal{Y}^{-1}$ amplification of the (non-rotated) fluctuation fields $\hat{\mathcal{U}}_3$ and $\hat{\mathcal{W}}_3$ is viscously regularized. 

To examine the dynamics within the CL, we introduce a rescaled fissure-normal coordinate $Y\equiv y/\tilde{\delta}$, where $\tilde{\delta}$ is the scaled thickness of the critical layer (and, again, $\tilde{\delta}/\tilde{\Delta}\to 0$ as $Re\to\infty$), and posit the following expansions:
\begin{eqnarray}\label{eq:CLscale}
u(x,y,z,t;Re)&\sim&\tilde{\Delta}\overline{U}_1(Y,z)+a'\frac{\tilde{\Delta}}{\tilde{\delta}}U'_2(x,Y,z)+\cdots,\label{CLexpansion:u}\\
v(x,y,z,t;Re)&\sim&\overline{a}\tilde{\delta}\,\overline{V}_4(Y,z)+a'V'_3(x,Y,z)+\cdots,\\
w(x,y,z,t;Re)&\sim&\overline{a}\overline{W}_2(z)+a'\frac{\tilde{\Delta}}{\tilde{\delta}}W'_2(x,Y,z)+\overline{a}\tilde{\Delta}\overline{W}_3(Y,z)+\cdots,\label{CLexpansion:w}\\
p(x,y,z,t;Re)&\sim&\overline{a}^2\overline{P}_4(z)+a'P'_3(x,Y,z)+\cdots.
\end{eqnarray}
For smooth matching with the streak and roll flow in the VF, the streak velocity becomes small, i.e. $\mathit{O}(\tilde{\Delta})$, while the mean spanwise velocity component remains $\mathit{O}(\bar{a})$ within the CL.  The latter scaling follows because, within the CL, $\bar{w}$ is not sheared at leading order.  In addition, the fissure-normal fluctuation velocity component $v'$ remains $\mathit{O}(a')$ as the CL is approached while the tangential components $u'$ and $w'$ blow-up algebraically; this growth is accounted for by the amplification of these fluctuation velocity components by 
the factor $\tilde{\Delta}/\tilde{\delta}$ in (\ref{CLexpansion:u}) and (\ref{CLexpansion:w}).

The requirement that viscous forces arise at leading order in the fluctuation equations yields the scaling relationship $Re\tilde{\delta}^3 =  \tilde{\Delta}^2$.  For the CL at $y=0$, the leading-order fluctuation equations thus reduce to
\begin{eqnarray}\label{eq:CLfluc}
	i\tilde{\alpha}\hat{U}_2\,+\,\partial_{Y}\hat{V}_3\,+\,i\tilde{\beta}\hat{W}_2&=&0,\label{eq:CLflucCONT}\\
	i\tilde{\alpha}\overline{U}_1\hat{U}_2\,+\,\hat{V}_3\partial_{Y}\overline{U}_1\,&=&\,-i\tilde{\alpha}\hat{P}_3+\bar{\Omega}_c\partial_Y^2\hat{U}_2,\label{eq:CLflucU}\\
	\partial_{Y}\hat{P}_3&=&0,\label{eq:CLflucV}\\
	i\tilde{\alpha}\overline{U}_1\hat{W}_2\,&=&\,-i\tilde{\beta}\hat{P}_3+\bar{\Omega}_c\partial_Y^2\hat{W}_2,\label{eq:CLflucW}
\end{eqnarray}
where the factor $\bar{\Omega}_c$ appears as a diffusion coefficient in the fluctuation momentum equations owing to the redefinition of the small parameter $\Delta\to\tilde{\Delta}$.  Since the leading-order mean streamwise-momentum equation requires $\partial_Y^2\overline{U}_1=0$, the streak velocity field varies linearly within the CL, i.e. $\overline{U}_1=\tilde{\lambda}(z)Y$, where the streak shear $\tilde{\lambda}(z)$ at $\mathcal{Y}=0$ is known from the solution of the Childress cell problem. Noting from (\ref{eq:CLflucV}) that $\hat{P}_3$ is independent of $Y$ and rearranging (\ref{eq:CLflucW}), \emph{viz.} 
\begin{equation}
\partial_{Y}^2\hat{W}_2-i\left(\frac{\tilde{\alpha}}{\bar{\Omega}_c} \tilde{\lambda}\right) Y\hat{W}_2 =\frac{i\tilde{\beta}}{\bar{\Omega}_c}\hat{\mathcal{P}}_3|_{\mathcal{Y}=0},
\end{equation}
$\hat{W}_2$ is found to satisfy a forced Airy-like equation, the solution of which may be found, e.g., in \cite{BalmforthJFM1997} and \cite{HallSherwinJFM2010}:
\begin{equation}\label{eq:CLwfluc}
\hat{W}_2 = -\bar{\Omega}_c^{-1/3}\left(\tilde{\alpha}\tilde{\lambda}\right)^{-2/3}\left(i\tilde{\beta}\hat{\mathcal{P}}_3|_{\mathcal{Y}=0}\right)\int_0^\infty e^{-i\left(\frac{\tilde{\alpha}\tilde{\lambda}}{\bar{\Omega}_c}\right)^{1/3}Y\varphi-\varphi^3/3}d\varphi.
\end{equation}
With $\hat{W}_2$ known, solutions for $\hat{U}_2$ and $\hat{V}_3$ also may be obtained, although these are not recorded here.

We next demonstrate that the nonlinear self-interaction of the fluctuation mode within the CL gives rise to a Reynolds stress divergence that ultimately drives a spanwise mean flow.  To determine this transverse mean-flow response, we apply the streamwise/fast-phase averaging operation to the $\bot$-momentum equations. When integrated across the CL, the leading-order (i.e. $\mathit{O}(\tilde{\Delta}^3)$) balance 
$\partial_Y^2\overline{W}_3-\partial_{Y}(\overline{V_3'W_2'})=0$
fails to induce a jump in the mean $x$-vorticity.  Accordingly, we turn to the mean equations at $\mathit{O}(\tilde{\Delta}^4)$:
\begin{eqnarray}
\partial_{Y}\tilde{P}_4 &=& 0,\label{eq:CL_Vmean}\\
\tilde{W}_2\partial_z\tilde{W}_2 &=&  -\partial_z\tilde{P}_4-\frac{\partial_z(\overline{W_2'W_2'})}{\overline{\Omega}_c^{2}} -\frac{\partial_{Y}(\overline{V_3'W_3'})}{\overline{\Omega}_c^{2}}+\partial_Y^2\tilde{W}_4,\label{eq:CL_Wmean}\\
\partial_{Y}\tilde{V}_4+\partial_z\tilde{W}_2 &=& 0.\label{eq:CL_cont}
\end{eqnarray}
Here, again, the mean transverse velocities and pressure have been rescaled so that $(\overline{V}_4,\overline{W}_2)=\bar{\Omega}_c(\tilde{V}_4,\tilde{W}_2)$ and $\overline{P}_4=\bar{\Omega}_c^2\tilde{P}_4$.
Balancing the fluctuation-induced forcing with mean diffusion in both the $\mathit{O}(\tilde{\Delta}^3)$ and the $\mathit{O}(\tilde{\Delta}^4)$ mean $\bot$-momentum equations requires the scaling relationship $(a')^2=\overline{a}/(\bar{\Omega}_c Re)$ to be satisfied.  Recalling that $\overline{a}=1/(\bar{\Omega_c}Re\tilde{\Delta}^2)$, the former relation is consistent with the requirement that $a'=\tilde{\Delta}\overline{a}$, an ordering already presumed in the analysis of the mean dynamics within the VF (see \S~\ref{sec:Childress}).  From the analysis of the VF it is also known that
\begin{eqnarray*}
-\partial_z{\tilde{\mathcal{P}}_4} = \tilde{\mathcal{W}}_2\partial_z\tilde{\mathcal{W}}_2.
\end{eqnarray*}
Since both $\tilde{P}_4$ and $\tilde{W}_2$ are independent of $Y$, this equality also must hold within the critical layer.  Subtracting this asymptotic balance from (\ref{eq:CL_Wmean}) and integrating the result across the CL (noting that the term involving the cross-correlation again integrates to zero) yields
\begin{equation}
\left[\partial_{Y}\tilde{W}_4\right]^+_- = \frac{1}{\bar{\Omega}_c^2}\int_{-\infty}^{\infty} \partial_z\left(\overline{W_2'W_2'}\right)\, dY,
\end{equation}
where $[\cdot]^+_-$ denotes the jump across the CL.  

Matching the mean $x$-vorticity at the edges of the CL with that at the center of the VF yields a crucial final scaling relationship $\tilde{\delta}=\tilde{\Delta}^2$. 
Using this relation in conjunction with the three other scaling relationships linking the asymptotic parameters $\tilde{\Delta}$, $\tilde{\delta}$, $\overline{a}$ and $a'$ gives
\begin{eqnarray}
\tilde{\Delta} = \left(\bar{\Omega}_c Re\right)^{-1/4}; \; \tilde{\delta} = \left(\bar{\Omega}_c Re\right)^{-1/2}; \; \overline{a} = \tilde{\Delta}^2; \; a' = \tilde{\Delta}^3;
\end{eqnarray}
implying $\Delta=Re^{-1/4}$ and $\delta = Re^{-1/2}$ (see table~1).  Moreover, the jump in the mean $x$-vorticity across the CL now can be expressed as
\begin{equation}\label{eq:CLwmean}
\left[\partial_{\mathcal{Y}}\tilde{\mathcal{W}}_3\right]^+_- = \frac{1}{\bar{\Omega}_c^2}\int_{-\infty}^{\infty} \partial_z\left(\overline{W_2'W_2'}\right)\, dY.
\end{equation}
Thus, as in VWI theory, a jump in the $x$-mean spanwise shear across the CL is induced by the nonlinear self-interaction of the Rayleigh mode within the CL. A key distinction, however, arises because of the separation in length scales between the rolls and the instability mode; in particular, a \emph{modulational} spanwise derivative of the Rayleigh-mode-induced Reynolds stress is operative in the present construction.

To evaluate this stress jump, we substitute (\ref{eq:CLwfluc}) into (\ref{eq:CLwmean}) to obtain 
\begin{equation}
\left[\partial_{\mathcal{Y}}\tilde{\mathcal{W}}_3\right]^+_- = \frac{4n_0 \mathcal{A}^2\tilde{\alpha}^{1/3}}{\bar{\Omega}_c^{7/3}}\partial_z\Bigg( \frac{A^2\tilde{\beta}^2\Big|\hat{\Pi}_3|_{\mathcal{Y}=0}\Big|^2}{\tilde{k}^2\tilde{\lambda}^{5/3}} \Bigg),
\end{equation}
where $n_0=2\pi(2/3)^{2/3}\Gamma(1/3)$ and $\Gamma$ is the gamma function \citep{HallSherwinJFM2010}.
Finally, normalizing so that $|\hat{\Pi}_3|_{\mathcal{Y}=0}=\tilde{\lambda}/\tilde{k}$ (conveniently yielding $\hat{\mathcal{V}}_3|_{\mathcal{Y}=0}=-i$) gives
\begin{equation}\label{eq:JumpNormalized}
\left[\partial_{\mathcal{Y}}\tilde{\mathcal{W}}_3\right]^+_- = \frac{4n_0 \mathcal{A}^2\tilde{\alpha}^{1/3}}{\bar{\Omega}_c^{7/3}}\partial_z\Bigg( \frac{A^2\tilde{\beta}^2\tilde{\lambda}^{1/3}}{\tilde{k}^4} \Bigg),
\end{equation}
where, again, for the fluctuation field $\partial_z$ is a modulational derivative.  
Consequently, the jump in spanwise shear stress across the CL is known in terms of the Rayleigh-mode parameters, i.e. the scalar amplitude $\mathcal{A}$, amplitude function $A(z)$, streamwise wavenumber $\tilde{\alpha}$ and spanwise wavenumber $\tilde{\beta}(z)$, and in terms of the streak shear stress $\tilde{\lambda}(z)$ and the to-be-determined roll vorticity $\bar{\Omega}_c$.  Note that $\bar{\Omega}_c$ also must be determined to relate the $\mathit{O}(1)$ streamwise wavenumbers $\tilde{\alpha}$ and $\breve{\alpha}$.  Since the latter of these wavenumbers is rendered dimensionless using only $Re^{1/4}$ and $l_y$, it is a more natural control parameter, e.g., for numerical computations of ECS at large but finite $Re$ using the full NS equations.

\subsection{Roll-flow energy balance}\label{subsec:energybalance}
To close the analysis, the nonlinear eigenvalue $\bar{\Omega}_c$ must be self-consistently determined.  The required calculation exploits the fact that, physically, the stress jump across the CL drives a spanwise flow within and, thence, a roll flow outside the VFs. More specifically, the work done by the mean viscous tangential stress at the CL/VF boundary is balanced by dissipation of roll kinetic energy within the UMZ.
This constraint can be utilized to determine the unknown roll vorticity $\bar{\Omega}_c$ by integrating the steady form of the $x$-mean $\perp$-kinetic energy equation over one roll cell, \emph{excluding} the critical layers at $y=0, 1$:
\begin{equation}
\begin{aligned}\label{eq:energetics}
\int_0^{L_z/2}&\int_{0^+}^{1^-}\Bigg[\overline{v}\partial_y\left(\frac{\overline{v}^2+\overline{w}^2}{2}\right)+\overline{w}\partial_z\left(\frac{\overline{v}^2+\overline{w}^2}{2}\right) +\overline{v}\partial_y\overline{p}+\overline{w}\partial_z\overline{p}\Bigg]dydz\\
&=-\int_0^{L_z/2}\int_{0^+}^{1^-}\bigg[\overline{v}\partial_y(\overline{v'v'})+\overline{v}\partial_z(\overline{v'w'})+\overline{w}\partial_y(\overline{v'w'})+\overline{w}\partial_z(\overline{w'w'})\bigg]dydz\\
&\;+\;\frac{1}{Re}\int_0^{L_z/2}\int_{0^+}^{1^-}\bigg(\overline{v}\nabla_\perp^2\overline{v}+\overline{w}\nabla_\perp^2\overline{w}\bigg)dydz.
\end{aligned}
\end{equation}
The leading-order balance in this roll-flow energy budget arises at $\mathit{O}(\Delta^8)$ and, using the $x$-mean incompressibility condition and boundary and symmetry conditions around the roll-cell perimeter, can be shown to include only the term proportional to $1/Re$.  Following an integration by parts, this term can be split into a diffusive flux of $\bot$-kinetic energy across the edge of the CL and viscous dissipation of $\bot$-kinetic energy within the domain of integration.
Thus, accounting for the forcing arising from the CL near $y=0^+$ and that from $y=1^-$, (\ref{eq:energetics}) reduces to
\begin{eqnarray}
-2\int_0^{L_z/2}\overline{w}\partial_y\overline{w}|_{y=0^+}dz\sim\int_0^{L_z/2}\hspace{-0.02\linewidth}\int_{0^+}^{1^-}\bigg(|\nabla_\bot\overline{v}|^2+|\nabla_\bot\overline{w}|^2\bigg)dydz=\int_0^{L_z/2}\hspace{-0.02\linewidth}\int_{0^+}^{1^-}\bar{\Omega}^2dydz.\qquad{}\label{eq:intermediate_energy_balance}
\end{eqnarray}
The dissipation integral on the right-hand side of (\ref{eq:intermediate_energy_balance}) is dominated by the contribution from the interior of the integration domain, i.e. from the UMZ, since the mean $x$-vorticity is of the same asymptotic magnitude, $\mathit{O}(\bar{a})$, within the VFs and UMZ, but the area of the UMZ is $\mathit{O}(1)$ while that of the VFs is $\mathit{O}(\Delta)$. Within the UMZ, the mean $x$-vorticity is homogenized so that $\bar{\Omega}\sim \bar{a}\bar{\Omega}_c$.  Consequently, (\ref{eq:intermediate_energy_balance}) further simplifies to
\begin{eqnarray}
-2\int_0^{L_z/2}\tilde{w}_2|_{y=0}\partial_\mathcal{Y}\tilde{\mathcal{W}}_3|_{\mathcal{Y}=0^+}dz&\sim&\frac{L_z}{2}.\label{eq:KEbalance}
\end{eqnarray}
Finally, employing (\ref{eq:JumpNormalized}) to evaluate (\ref{eq:KEbalance}) yields
\begin{equation}\label{eq:OmegEqn}
\bar{\Omega}_c^{7/3} = \frac{32\mathcal{A}^2n_0\tilde{\alpha}^{1/3}}{\pi^2}\int_0^{L_z/2}\, \sum_{n=0}^{n=\infty} \Bigg[ \frac{1}{n^2}\tanh\left(\frac{n\pi}{L_z}\right) \sin\left(\frac{2n\pi z}{L_z}\right)\Bigg]\,\partial_z\left(\frac{A^2\tilde{\beta}^2\tilde{\lambda}^{1/3}}{\tilde{k}^4}\right)dz.
\end{equation}
This expression relates $\bar{\Omega}_c$ to known or computable quantities, thereby closing the problem for the construction of inertial ECS.

\section{Spatially-distributed inertial exact coherent states}\label{sec:ECS}
In this section, ECS are constructed using the results of the asymptotic analysis described in \S~3. For this purpose, the solution of the Childress cell problem is replaced by the large (but finite) $Re$ solution of the steady advection--diffusion equation (\ref{eq:UMZ_UmeanU}) for the leading-order streak velocity over the entire spatial domain. It proves more convenient to extract the full (2D) streak velocity field within the VF in this manner than to attempt to reconstruct this field (rather than just the streak shear $\tilde{\lambda}(z)$ at the VF centreline) from the formal solution to the Childress cell problem. Moreover, as noted in the discussion of figure \ref{fig:ShearCoreVorticity}, the former approach has the advantage of viscously regularizing the passive singularities in the solution of the Childress cell problem arising near the cell corners.  

The ECS we construct are spatially distributed in that the rolls and streaks are stacked in the $y$ direction, being embedded in unbounded plane Couette flow.  Because the Rayleigh instability modes are strictly localized within the spatially-segregated fissures (owing to the scale separation between the wavy instabilities and the roll/streak flow), the required stacking is simpler than that performed in \cite{HallJFM2018}.  In the latter investigation, a countable infinity of viscous VWI states (each related to the lower-branch equilibrium solution EQ7) is shown to be an asymptotic solution in unbounded plane Couette flow provided the $x$-varying fluctuation pressure satisfies a certain \emph{global} elliptic equation.  In contrast, the fluctuation pressure field associated with the spatially-distributed inertial ECS constructed here decays exponentially away from the centre of each VF.  The only subtlety is that although roll and streak flow in each layer is steady, the Rayleigh instability mode is stationary only for the VF at $y=0$.  More generally, the instability mode localized within the VF at $y=n$, for integer $n$, propagates with a phase speed $c_n=n$, i.e. the average of the homogenized streamwise speeds of the adjacent upper and lower UMZs (\emph{viz.} $[(n+1/2)+(n-1/2)]/2$).  Nevertheless, the homogenized vorticity $\bar{\Omega}_c$ within each UMZ has the same value, implying the rolls have the same circulation strength independently of their $y$ location.  This deduction, however, applies only to unbounded plane Couette flow and does not hold for base flow profiles that vary nonlinearly with $y$.  {\cb{The ECS solution algorithm is summarized in appendix~\ref{app:ECSsoln}.}}



\subsection{Three-dimensional structure}\label{subsec:RECONSTRUCTIONS}
\begin{figure}
	\begin{center}
		\includegraphics[scale=0.495]{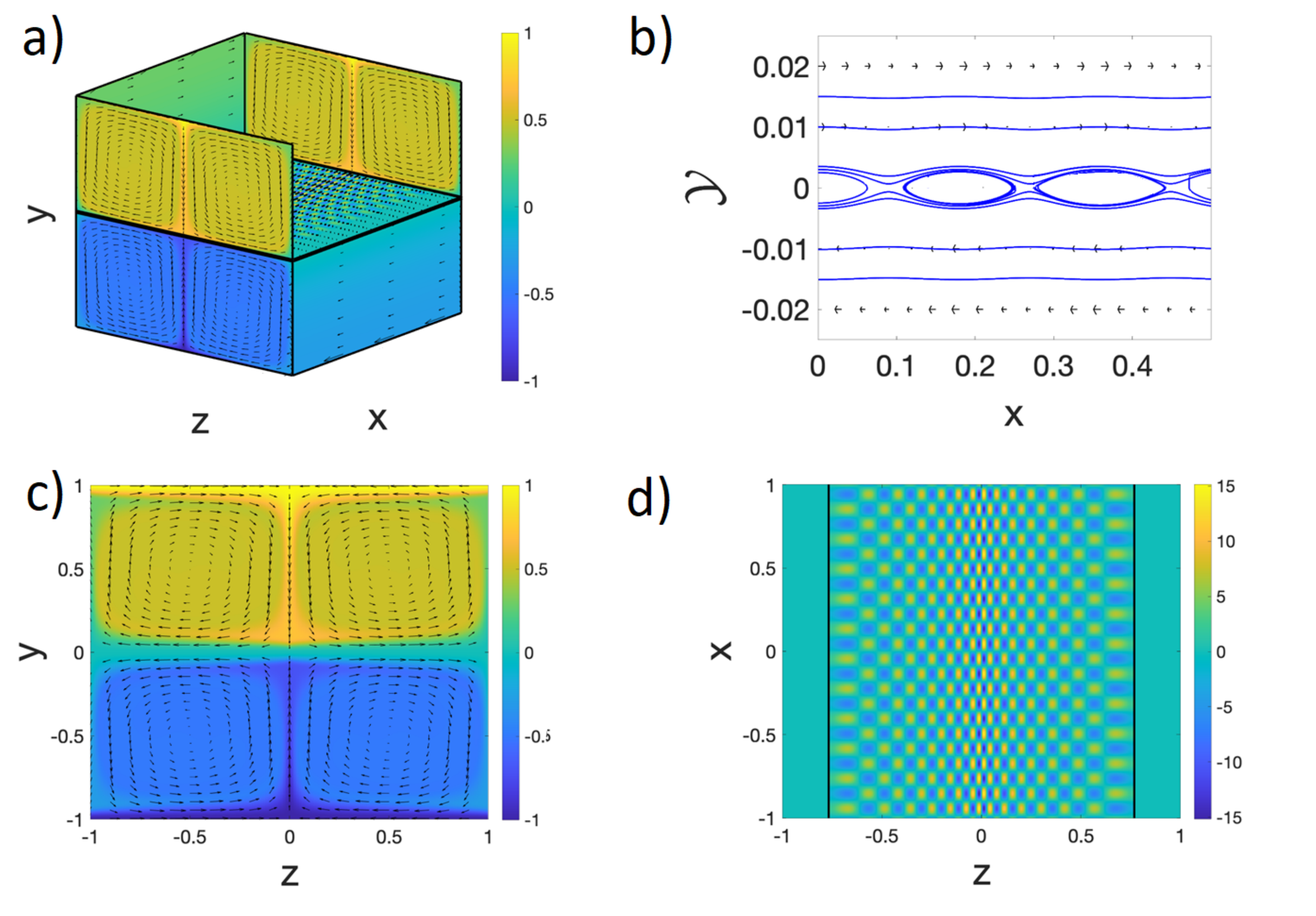}
		\caption{3D reconstruction of inertial ECS exhibiting VFs coupled to adjacent UMZs in unbounded plane Couette flow. Input parameters are {\cb{$\mathcal{A}=5$, $\breve{\alpha}=0.576$}} and $L_z = 2$.  ({\cb{$Re = 1.36 \times 10^7$}} for computation of the streak velocity field.) (a) 3D rendering. Color indicates the streamwise-averaged streamwise flow speed in the $y$--$z$ and $x$--$y$ planes and the fissure-normal fluctuation velocity, scaled by $O({\Delta}^3)$, in the exposed $x$--$z$ plane.  Arrows show in-plane velocity vectors.  (b) Close-up view of streamlines in {\cb{a streamwise/wall-normal plane (i.e. $z=-0.0437$) on which the spanwise fluctuation velocity vanishes, and within the VF centred on $y=0$, highlighting}} the appearance of the fine-scale Kelvin's cat's-eyes vortex pattern within the VF.  (Arrows are also plotted to indicate in-plane velocity vectors.) (c) Mean velocity field in the fissure-normal/spanwise-plane. Color indicates the value of the streamwise-averaged streamwise velocity component while arrows show the roll velocity field.  (d) Fissure-normal fluctuation velocity component in a streamwise/spanwise plane within the VF near $y=0$. Color indicates the value of this component normalized by ${\Delta}^3$.  The solid black lines indicate caustics, which partition the plane into an illuminated region ($-0.75\lesssim z\lesssim 0.75$) and shadow zones ($|z|\gtrsim 0.75$). }
		\label{fig:Reconstruction}
	\end{center}
\end{figure}

Figure \ref{fig:Reconstruction}(a) shows a ``unit ECS'' excised from the flow domain centred on the VF at $y=0$ {\cb{and intended to mimic the highlighted region in figure~\ref{fig:mean_force_balance}(b)}}. Color indicates the  streamwise-averaged streamwise flow speed in the $y$--$z$ and $x$--$y$ planes, as well as the $O({\Delta}^3)$ fissure-normal fluctuation velocity in the exposed $x$--$z$ plane. The parameters used to generate this ECS are {\cb{$\mathcal{A}=5$, $\tilde{\alpha}=0.35$}} and $L_z = 2$. With these inputs, the $x$-mean vorticity {\cb{$\bar{\Omega}_c = 7.35$}} is self-consistently computed. The true streamwise wavenumber, i.e. the dimensional streamwise wavenumber divided by $Re^{1/4}l_y^{-1}$ independently of $\bar{\Omega}_c$, is found to be {\cb{$\breve{\alpha} = 0.576$}}.  For ease of interpretation, visualizations are presented in three planes: the $y$--$z$ plane at $x=0$~(c), an $x$--$z$ plane within the VF~(d) and {\cb{an $x$--$y$ plane restricted in $y$ to show a close-up near the center of the VF~(b). }}

Figure \ref{fig:Reconstruction}(c) shows a plane perpendicular to the streamwise flow. Only mean velocities are plotted because the fluctuation fields are exponentially small except within the asymptotically thin VF. The colors indicate the values of the mean streamwise velocity component, while the arrows show the in-plane roll velocity field with the size of the arrows indicating the flow velocities relative to an arbitrary maximum value. This construction enables the rolls to be visualized but it should be recalled that, asymptotically, the roll flow is weak relative to the streamwise velocity component.
The steady solution clearly exhibits the homogenization driven by the rolls
as well as the spontaneous emergence of a fissure at $y = 0$. Additional shear layers located above at $y=1$ and below at $y=-1$ also are apparent. As expected from symmetry considerations, the streak speed within the upper UMZ is homogenized to a value $\overline{u}_0 = 1/2$ while the lower UMZ has $\overline{u}_0 = -1/2$.  For the upper UMZ, fluid having relatively high (low) streamwise momentum is carried downward (upward) in narrow jets centred on $z=0$ ($z=\pm 1$).

Figure \ref{fig:Reconstruction}(d) shows the fissure-normal fluctuation velocity component $\mathcal{V}_3'$ in a streamwise/spanwise plane located within the VF  close to the CL (i.e. close to $\mathcal{Y}=0$). Color is used to represent flow speeds with yellow corresponding to positive and blue to negative values. This visualization reveals two important features of the inertial ECS supported by the SSP investigated here. Firstly, caustics partition the spanwise domain within the VF into an illuminated region, within which the fluctuations are confined, and shadow zones, where the instability mode is exponentially small. Note that only the fluctuations are plotted in figure \ref{fig:Reconstruction}(d); the mean fields do not vanish in the shadow regions, so the fluid is in motion there. The shadow zones are observed to align with the upwelling (downwelling) regions in the mean flow for $y>0$ ($y<0$), suggesting that the caustics arise from divergent straining of the shear layer to the extent that the shear within the VF is too diffuse to support the Rayleigh instability mode. A second feature that is clearly evident is the slow $z$ modulation of the Rayleigh mode spanwise wavenumber and amplitude; indeed, the lines of constant instability-wave phase appear to bend in this plane, forming cusps at the caustics.  This patterning highlights the horizontal scale separation between the instability mode and the roll/streak flow.

Figure \ref{fig:Reconstruction}(b) shows a close-up view of streamlines in {\cb{a streamwise}}/fissure-normal plane within the VF centred on $y=0$. {\cb{On this particular symmetry plane, the spanwise fluctuation velocity vanishes, thereby highlighting}} the emergence of a fine-scale Kelvin's cat's-eyes vortex pattern within the VF.  This patterning is commonly observed in the neighborhood of critical layers when streamlines are plotted in a reference frame traveling with the marginally stable disturbance mode.

\subsection{Parametric dependencies}\label{subsec:PARAMSTUDY}
We next analyze the impact of varying the key control parameters $\mathcal{A}$, $\breve{\alpha}$ and $L_z$ on the ECS. For this purpose, a convenient metric is the homogenized value of the $x$-mean $x$-vorticity $\bar{\Omega}_c$ within the UMZs.
Equation (\ref{eq:OmegEqn}) gives $\bar{\Omega}_c$ as a function of $L_z$, $\tilde{\alpha}$ and $\mathcal{A}$.  To recover the dependence of $\bar{\Omega}_c$ on the true streamwise wavenumber $\breve{\alpha}$, the substitution $\breve{\alpha}=\left[\bar{\Omega}_c(\tilde{\alpha})\right]^{1/4}\tilde{\alpha}$ is made.  In addition, for a given roll-cell wavelength $L_z$, the dependence of $\bar{\Omega}_c$ on $\mathcal{A}$ can be removed by scaling $\bar{\Omega}_c\to\mathcal{A}^{6/7}\bar{\Omega}_c$, provided that the true streamwise wavenumber is also rescaled according to $\breve{\alpha}\to \mathcal{A}^{3/14}\breve{\alpha}$.
\begin{figure}
	\hspace{-0.05\linewidth}\includegraphics[scale=0.58]{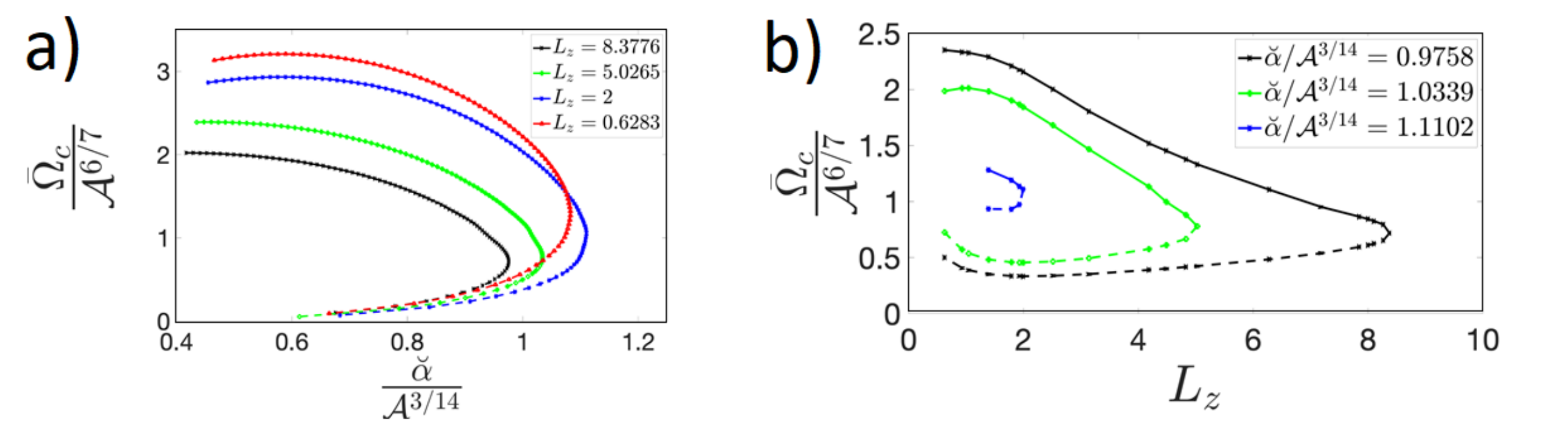}
	\caption{The homogenized value of the roll vorticity $\bar{\Omega}_c$ normalized by $\mathcal{A}^{6/7}$ plotted (a)~versus the ``true'' streamwise wavenumber $\breve{\alpha}$ normalized by $\mathcal{A}^{3/14}$, with the different curves corresponding to different constant values of the roll wavelength $L_z$; and (b) versus $L_z$, with the different curves corresponding to different constant values of $\breve{\alpha}$.}\label{fig:Param}
\end{figure}
When rescaled in this way, plots of $\bar{\Omega}_c$ versus $\breve{\alpha}$ for fixed $L_z$ but varying $\mathcal{A}$ will collapse onto a single curve.  Physically, the roll circulation achieved for a given $L_z$, $\mathcal{A}$ and $\breve{\alpha}$ also can be realized by appropriately modifying both $\mathcal{A}$ and $\breve{\alpha}$.



Figure \ref{fig:Param}(a) shows the $x$-mean vorticity versus the true streamwise wavenumber $\breve{\alpha}$ for different given values of $L_z$. {\cb{When}} $\bar{\Omega}_c$ is plotted versus the true streamwise wavenumber $\breve{\alpha}$, a saddle-node bifurcation with upper- and lower-branch solutions is revealed{\cb{; in contrast, as a function of $\tilde{\alpha}$, $\bar{\Omega}_c$ is not multi-valued and decreases monotonically to zero as $\tilde{\alpha}$ is increased beyond a threshold value. This latter trend may be understood by recalling the Eikonal equation $\tilde{\alpha}^2+\tilde{\beta}^2(z)=\tilde{k}^2(z)$. For a given $L_z$, the $z$-varying profile of the eigenvalue $\tilde{k}^2$ qualitatively resembles that of the streak shear $\tilde{\lambda}(z)$, which itself is \emph{independent} of $\tilde{\alpha}$, shown in figure~\ref{fig:ShearCoreVorticity}:  the total horizontal wavenumber of the neutral Rayleigh mode is larger where the shear is locally larger and the fissure is locally thinner (and conversely).  As $\tilde{\alpha}$ is increased, the Eikonal equation therefore requires the width of the illuminated region to decrease.  Consequently, both the magnitude of the $z$-integrated forcing from the Rayleigh mode and the resulting value of $\bar{\Omega}_c$ also must decrease. The caveat is that, as the turning point is approached from the upper branch, our ECS solutions become non-uniformly asymptotic; that is, for the lower-branch states, ever larger values of $Re$ are required to ensure that the width of the illuminated region remains asymptotically larger than the widths of the caustics and the streamwise jets. For this reason, we have used a dashed line to designate the lower-branch states in figure~\ref{fig:Param}(a,b).}}

For the upper-branch solutions {\cb{in figure~\ref{fig:Param}(a)}}, it can be seen that $\bar{\Omega}_c$ increases monotonically as the roll-cell size is decreased.  In contrast, the roll vorticity associated with the lower-branch states does not vary monotonically with $L_z$.  Moreover, the bifurcation curve for each fixed $L_z$ possesses a maximum along the upper branch, implying that for a given $L_z$ there exists a specific streamwise wavenumber at which the roll-cell circulation is maximized. Figure \ref{fig:Param}(b) shows the $x$-mean vorticity as a function of cell width for a range of given, constant streamwise wavenumbers. In this figure, too, the solution bifurcates into upper and lower branches.  The upper branch is characterized by stronger roll-cell circulation arising at smaller $\breve{\alpha}$ (i.e. for streamwise-varying instability modes with longer wavelengths), while the lower branch exhibits the opposite trend. 
{\cb{Again, the}} primary qualitative distinction between the upper- and lower-branch solutions is that the illuminated regions associated with the upper-branch states are larger than those associated with the corresponding lower-branch ECS.

\section{Conclusion}\label{sec:CONCLUSION}
In turbulent wall flows at large Reynolds number, the ensemble or long-time mean viscous force is sub-dominant outboard of the peak in the Reynolds stress and hence over the majority of the turbulent flow domain. Mounting evidence indicates that the instantaneous streamwise velocity in this inertial domain is characterized by regions of quasi-uniform momentum (UMZs) separated by spatially-segregated internal shear layers (i.e. vortical fissures).  In this investigation, a first-principles self-sustaining process theory has been derived from the incompressible Navier--Stokes equations in the large Reynolds number limit that can account for key attributes of the resulting staircase-like profiles of streamwise velocity.  Chief among these attributes is that the suitably normalized fissure thickness decreases as the friction Reynolds number $Re_\tau=u_\tau h/\nu$ increases; that the dimensional jump in flow speed across each VF scales with the friction velocity $u_\tau$; and that the volume-mean viscous force is, in fact, sub-dominant in this dynamical process.

\begin{figure}
	\begin{center}
				\includegraphics[width=1.1\linewidth, angle=0]{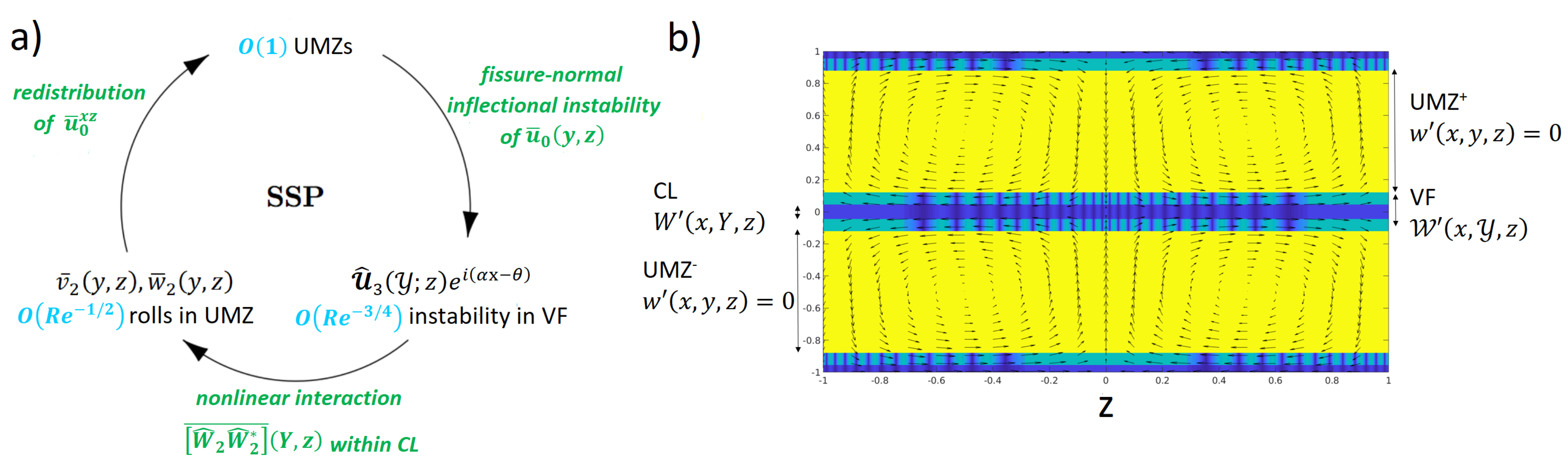}
		\caption{Schematic diagram of a mechanistic self-sustaining process for UMZs and VFs in the inertial domain of turbulent wall flows.  The feedback loop shown in~(a) indicates that rolls having $\mathit{O}(Re^{-1/2})$ circulation strength redistribute the background shear flow to induce an $\mathit{O}(1)$ inflected streamwise flow.  As depicted in (b), the counter-rotating and stacked rolls are sufficiently strong to differentially homogenize the background flow, thereby creating and maintaining both UMZs (highlighted in yellow) and internal shear layers (VFs, indicated in blue).  The wall-normal ($y$) inflections in the streamwise-mean streamwise velocity support an $\mathit{O}(Re^{-3/4})$ Rayleigh (inviscid shear) instability mode that has a streamwise ($x$) wavelength $2\pi/{\alpha}\ll 1$ that is commensurate with the VF thickness.  Consequently, the Rayleigh mode is confined to the VF, where it is refracted and rendered three-dimensional by the comparably slow spanwise variation in the fissure thickness (not depicted), and the resulting ECS are inherently multiscale.  The 3D Rayleigh mode suffers a critical-layer singularity, causing the magnitude of the $x$-varying spanwise velocity component to be amplified to $\mathit{O}(Re^{-1/2})$. The resulting nonlinear self-interaction of the Rayleigh mode within the CL drives the roll motions in the UMZs through a modulational (i.e. slow spanwise) Reynolds stress divergence involving this velocity component.
		\textcolor{white}{Test test, why does this need to be on another line?}}
		\label{fig:newSSP}
	\end{center}
\end{figure}

Figure~\ref{fig:newSSP} depicts the key components of the proposed inertial-layer SSP.  As in vortex--wave interaction (VWI) theory, streamwise rolls induce $\mathit{O}(1)$ streamwise streaks through the lift-up mechanism.  A crucial distinction, however, is that in the inertial domain the comparably weak rolls must nevertheless have a circulation strength that is asymptotically larger than $\mathit{O}(1/Re)$ to ensure that these large-scale roll and streak components of the turbulence are not (in a volume-averaged sense) dynamically influenced by viscous forces, unlike ECS solutions of the VWI equations.  Indeed, in the present inertial-layer SSP theory, we find that the roll strength is $\mathit{O}(Re^{-1/2})$.  This sub-$Re^{-1}$ decay is also a necessary condition for stacked arrays of counter-rotating roll vortices to  differentially homogenize the background shear flow and thereby generate slender embedded internal layers of high vorticity.  Each internal layer (VF) has a characteristic thickness that decreases with $Re$.  In contrast, the lower- and upper-branch equilibrium solutions EQ7 and EQ8 in plane Couette flow, which share the same {\cb{stacked}} roll configuration as the ECS constructed here, remain viscously dominated even as $Re\to\infty$.  Consequently, neither EQ7 nor EQ8 exhibits thinning internal shear layers or genuinely homogenized zones of streamwise momentum.

Apparent discontinuities in the inertial-ECS streak velocity and roll vorticity are smoothed by viscous forces and torques acting within the VFs and narrow streamwise jets demarcating the boundary of each roll cell.  The inflectional streak shear within each fissure supports a neutral Rayleigh instability (or ``fluctuation'') mode having an $\mathit{O}(Re^{-3/4})$ characteristic size and a streamwise wavelength that scales with the VF thickness.  Thus, 
inertial ECS are inherently multiscale: the streamwise-varying fluctuation fields and the streamwise-mean roll/streak flow exhibit disparate spatial scales, a feature that is  accommodated in the theory via a WKBJ representation of the instability mode.  Moreover, the fluctuation fields are exponentially localized within the fissures, since that is where the shear is confined.  The Rayleigh mode, which might be expected to vary primarily in the streamwise and fissure-normal directions, is refracted in the spanwise direction and rendered 3D by the comparably slow spanwise variation in the thickness of each fissure.  This three-dimensionality is essential because, at least in the present formulation, the fluctuation-induced Reynolds stresses associated with the inviscid marginal mode necessarily vanish within the VF; thus, the roll motions are not directly driven there.  

Instead, a critical layer (CL) mechanism is operative.  More specifically, although the (appropriately rotated) two-component, 2D marginal Rayleigh mode is a smooth function of the fissure-normal coordinate, the full three-component, 3D marginal mode exhibits a CL singularity at the centre of each VF.  As in VWI theory, the resulting amplification of the tangential fluctuation velocity components is regularized by viscous forces acting within the CL.  (It is conceivable that nonlinear regularization also may be realizable, although that possibility is not pursued here.)   The nonlinear self-interaction of the Rayleigh mode within the CL induces a jump in the $x$-mean spanwise shear or, equivalently, the $x$-mean streamwise vorticity across the CL via a \emph{modulational} (i.e. slow spanwise) divergence of the spanwise Reynolds stress component.  In turn, this shear drives a tangential mean flow within -- and ultimately a roll flow outside of -- the VF in which the CL is embedded.  (Since this flow is transverse to the streamwise velocity and driven by spanwise stress gradients, it may be interpreted as a \emph{Prandtl secondary flow of the second kind}.) In steady state, the work done by the fluctuation-induced mean viscous tangential stress at the edges of the CL is balanced by viscous dissipation of roll kinetic energy within the adjacent UMZs, thereby closing the inertial-domain SSP.  

Visualizations of the inertial ECS clearly show the slow spanwise modulation of the Rayleigh-mode spanwise wavenumber as well as the emergence of caustics, which separate the VFs into illuminated and shadow regions.  Within these zones, the fluctuations have finite and exponentially small amplitudes, respectively.  The caustics arise because, at specific spanwise locations, the roll flow strains the embedded VF to the extent that the streak shear is too diffuse to continue to support the marginal Rayleigh mode.  As is common in shear flows, the ECS we identify exhibit saddle-node bifurcations, with the primary distinction between the upper- and lower-branch solutions being the magnitude of the roll-cell circulation and the width of the illuminated regions (both are larger for the upper-branch states).  

\begin{figure}
	\begin{center}
		\includegraphics[width=0.95\linewidth]{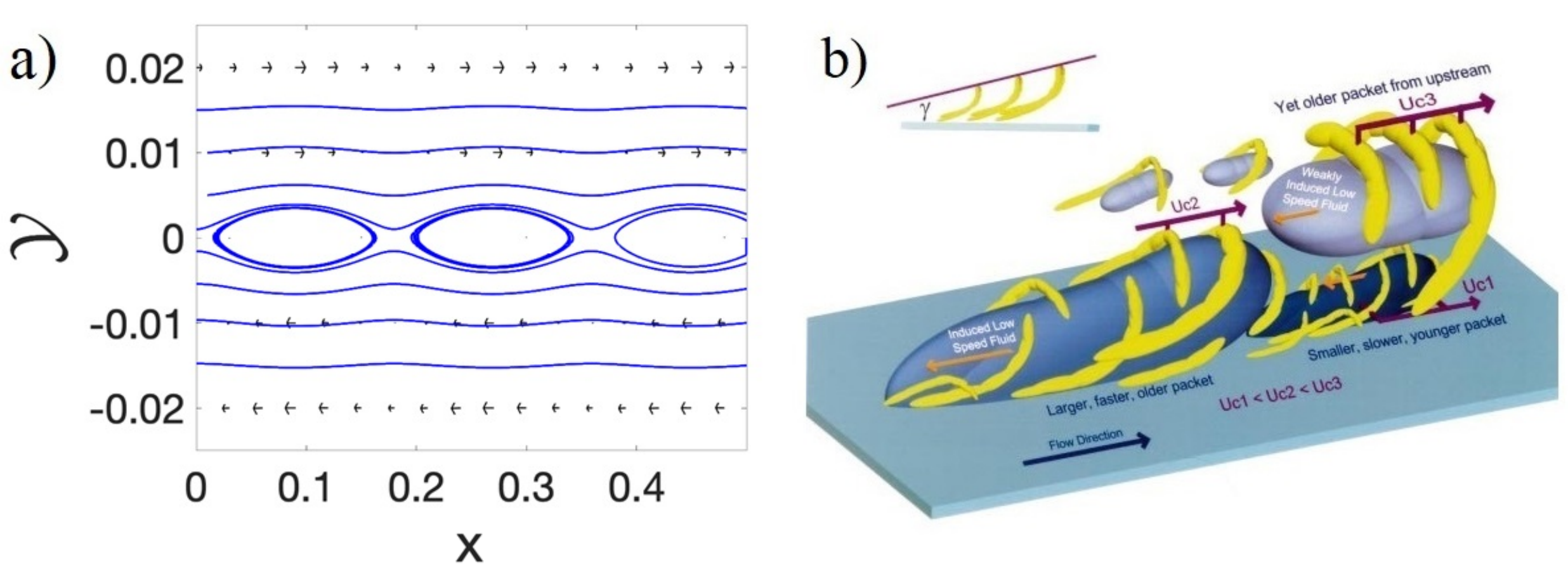}
		\caption{(a) Streamlines within the VF associated with an inertial-ECS solution computed for {\cb{$\mathcal{A}=5$, $\breve{\alpha}=0.576$}} and $L_z = 2$ (with {\cb{$Re = 1.36 \text{x} 10^7$}}) showing a fine-scale Kelvin's cat's-eyes vortex pattern in the {\cb{streamwise}}/wall-normal plane {\cb{$z=-0.6738$ on which the spanwise fluctuation velocity vanishes (cf. figure~\ref{fig:Reconstruction}(b), showing the corresponding cat's-eyes pattern in the plane $z=-0.0437$).}}  (b)~Schematic taken from \cite{AdrianJFM2000} showing fine-scale hairpin and ``cane'' vortices (in yellow) separating boluses of large-scale streamwise flow having quasi-uniform momentum (in grey).}  
		\label{fig:XY}
	\end{center}
\end{figure}

In a co-moving reference frame, the streamlines within each fissure adopt a characteristic Kelvin's cat's-eyes vortex pattern.  Intriguingly, this streamline pattern seems to accord  qualitatively with the conceptual model of \cite{AdrianJFM2000}, who proposed that regions of high shear separating the UMZs comprise the heads of hairpin and/or ``cane'' vortices, as shown schematically in figure~\ref{fig:XY}(b).  Regardless of the relevance or not of hairpin vortices \emph{per se}, flow visualizations by \cite{AdrianJFM2000} and others  reveal small-scale rotary motions, consistent with the structure of the ECS supported by the proposed SSP, as illustrated in figure \ref{fig:XY}(a).  Perhaps more significantly, the schematic shown in figure~\ref{fig:XY}(b) suggests a disparity between  the scale (diameter) of these rotary motions and the transverse scale of the boluses of fluid having uniform momentum, again in apparent qualitative agreement with the multiscale structure intrinsic to the proposed SSP.  In particular, the rotary motions in the inertial ECS have a size that is commensurate with the $\mathit{O}(\Delta)$ thickness of the fissure and therefore much smaller than separation distance between adjacent VFs.

Of course, there are certain evident limitations on the potential applicability of the SSP theory developed here to turbulent wall flows.  Firstly, the predicted VF thickness scales as $Re^{-1/4}$, while data from DNS and laboratory experiments indicate that the thickness of a ``representative" fissure scales in proportion to $Re_\tau^{-1/2}$.  It seems conceivable, however, that the difference between the predicted and measured scalings may be at least partly attributable to the different definitions of the Reynolds numbers used.  In particular, the length scale used in the definition of $Re$ is the separation distance $l_y$ between the VFs, while the length scale $h$ used to define $Re_\tau$ is the boundary-layer height or channel half-height.  Moreover, for turbulent wall flows, 
$l_y$ varies with the wall-normal coordinate, and the ratio of the characteristic separation distance $\langle l_y\rangle$ to $h$ varies with $Re_\tau$.  In appendix~\ref{app:VFscaling}, we show that, by accounting for this variation in the context of the self-similar layer hierarchy admitted by the mean momentum balance (e.g. \cite{KlewickiJFM2013,KlewickiIUTAM2013}), the associated characteristic dimensional VF thickness $\Delta_f$ normalized by $h$ scales in proportion to $Re_\tau^{-7/16}$, i.e. in much closer agreement with empirical estimates.

A second, related restriction is that the ECS we construct are asymptotic solutions of the Navier--Stokes equations only for the non-physically-realizable case of unbounded Couette flow.  Nevertheless, our ultimate aim is to systematically extend the theory to treat inertial ECS arising in flows (e.g. plane Poiseuille flow) having velocity profiles that vary nonlinearly with the wall-normal coordinate.  Preliminary considerations indicate that for solutions to exist the VFs necessarily will be non-uniformly spaced, in accord with observations of turbulent wall flows.  In addition, it would be desirable to develop a time-dependent reduced PDE model of turbulence in the inertial layer by leveraging the most robust features of the asymptotic SSP identified in this study (a general strategy advocated, e.g., by \cite{ChiniJFMFoF2016}); for example, by exploiting the confinement of the fluctuations to the fissures and the \emph{quasilinear} mathematical structure of the asymptotically-simplified system.  This extension would enable an even tighter link to be made with the 1D UMZ/VF turbulence model recently developed by \cite{BautistaJFM2018}, e.g. by allowing for time-dependent motion (``wafting'') of the fissures, while also placing the 1D model on a more secure theoretical footing.  Finally, the asymptotic ECS computed here for a given set of parameters presumably could be used as a very good initial iterate in a Newton search for a finite (but large) Reynolds number realization of this ECS employing the full Navier--Stokes equations.  

In summary, a chief merit of the new SSP identified here is that it highlights the distinction between viscous and inertial ECS.  The SSP also provides a plausible mechanism, derived directly from the NS equations, for the observed UMZ/VF profiles of streamwise velocity.  In this mechanism, the VFs play a dynamically active role in the sustenance and persistence of the UMZs.  
Perhaps most significantly, the ECS supported by the new SSP provide a concrete dynamical realization of the conceptual model of the singular
nature of turbulent wall flows proposed by \cite{KlewickiJFM2013,KlewickiIUTAM2013}, lending further support to the ``boundary-layers-within-\emph{the}-turbulent-boundary-layer" paradigm.

\newpage
\noindent
\textbf{Acknowledgments.}  The authors are pleased to acknowledge Dr. A. Ebadi for collecting and plotting the data shown in figure~1(b,c) and Professor J. Gibson for generating figure~4(b).  The authors are also grateful for financial support from the National Science Foundation through grant no. NSF CBET-1437851, the Office of Naval Research through grant no. N000141712307 and the Australian Research Council through grant no.  DP150102593.  Finally, GPC would like to acknowledge the KITP Program on \emph{Planetary Boundary Layers}, supported in part by the National Science Foundation under grant no. NSF PHY-1748958, and the Woods Hole Summer Program in Geophysical Fluid Dynamics, where parts of this work were completed in 2018.\\

\noindent
\textbf{Declaration of interests.} The authors report no conflict of interest.

\appendix
{\cb{
\section{Analysis of Childress cell problem}\label{app:ChildressDetails}
Within the connected domain shaded in light grey in figure~\ref{fig:CHILDRESS}(b), $\overline{\mathcal{U}}_0$ is governed by the single equation
\begin{equation}\label{eq:Childress}
\tilde{\mathcal{V}}_N\partial_\mathcal{N}\overline{\mathcal{U}}_0 + \tilde{\mathcal{V}}_s\partial_s\overline{\mathcal{U}}_0 = \partial_\mathcal{N}^2\overline{\mathcal{U}}_0.
\end{equation}
Here, $\mathcal{N}$ and $s$ are coordinates normal to and tangent to the cell boundary; within the VF centred at $y=0$, for example, these coordinates are equal to $\mathcal{Y}$ and $z$, respectively. The required suturing exploits the fact that the scalar field is passively advected through each corner region. 

By making a suitable change of variables, $\overline{\mathcal{U}}_0(s,\mathcal{N})= \overline{\mathrm{U}}(\zeta,\eta) $, with the so-called Crocco or Von Mises coordinates
\begin{equation}\label{eq:Crocco}
\zeta = \int_0^s\tilde{\mathcal{V}}_s(\hat{s})d\hat{s}, \quad \eta = \frac{1}{2}\int_0^\mathcal{N}\tilde{\mathcal{V}}_s(s)d\hat{\mathcal{N}},
\end{equation}
the advection-diffusion equation (\ref{eq:Childress}) can be transformed into the diffusion equation
\begin{equation}\label{eq:Childress2}
\partial_\zeta\overline{\mathrm{U}}=\frac{1}{4}\partial_\eta^2\overline{\mathrm{U}}.
\end{equation}
Equation (\ref{eq:Childress2}) is solved subject to the boundary conditions
\begin{eqnarray}
\overline{\mathrm{U}}(\zeta,0) &=& 0~\text{on}~0<\zeta<l_1,\\
\partial_\eta\overline{\mathrm{U}}(\zeta,0) &=& 0~\text{on}~l_1<\zeta<l_2,\\
\overline{\mathrm{U}}(\zeta,0) &=& 1~\text{on}~l_2<\zeta<(2l_1+l_2),\\
\partial_\eta\overline{\mathrm{U}}(\zeta,0) &=& 0~\text{on}~(2l_1+l_2)<\zeta <L \equiv 2(l_1+l_2),
\end{eqnarray}
where $l_1$ and $l_2$ are the lengths of the fissure-parallel and fissure-normal cell edges, respectively, measured in the coordinate $\zeta$, and subject to the periodicity condition
\begin{equation}
\overline{\mathrm{U}}(\zeta + nL,\eta) = \overline{\mathrm{U}}(\zeta,\eta),\qquad n = 1,2,3,\cdots
\end{equation}
in the time-like coordinate $\zeta$. As first shown by \cite{JimenezJFM1987}, exploiting this periodicity requirement enables the formal solution of the Childress cell problem to be expressed as
\begin{equation}\label{eq:ChildIntegral}
\overline{\mathrm{U}}(\zeta,\eta) = \frac{\eta}{\sqrt{\pi}}\int_0^\infty\frac{\mathrm{U}_w(\zeta-p)}{p^{3/2}}e^{-\eta^2/p}dp.
\end{equation}
An integral equation for the unknown function $\mathrm{U}_w(\zeta)\equiv\overline{\mathrm{U}}(\zeta,\eta=0)$ arising in the integrand of (\ref{eq:ChildIntegral}) is obtained by setting $\partial_\eta\overline{\mathrm{U}}(0,\eta) = 0$ on the two fissure-normal cell edges. Accurate numerical solution of the resulting integral equation requires careful treatment of the singular behavior of $\overline{\mathrm{U}}$ in the corner regions and benefits from specialized acceleration techniques to evaluate numerous slowly-converging infinite summations; see \cite{ChiniPoF2009} for details.}}

\section{Caustics}\label{app:Caustic}
The WKBJ solution for the fluctuation pressure field breaks down at the caustics. To find a uniformly valid solution for all $z$ within the internal shear layers (VFs), we employ an ansatz similar to that made by \cite{LudwigCPAM1966} and  \cite{MckeePCPS1973}:
\begin{equation}
\mathcal{P}' \sim \tilde{\Delta}^{17/6}\hat{\mathcal{P}}_3(\mathcal{Y},z)\left[F(\mathcal{Y},z)\text{Ai}\left(-\frac{\tilde{\xi}(z)}{\tilde{\Delta}^{2/3}}\right)+i\tilde{\Delta}^{1/3}G(\mathcal{Y},z)\text{Ai}'\left(-\frac{\tilde{\xi}(z)}{\tilde{\Delta}^{2/3}}\right)\right]e^{i(\tilde{\alpha}/\tilde{\Delta})(x-ct)}+c.c.,
\end{equation}
where here $\hat{\mathcal{P}}_3$ is \emph{defined} to satisfy 
\begin{equation}
    \partial_\mathcal{Y}^2\hat{\mathcal{P}}_3-2\frac{\partial_\mathcal{Y}\overline{\mathcal{U}}_0}{\overline{\mathcal{U}}_0-c}\partial_\mathcal{Y}\hat{\mathcal{P}}_3=\tilde{k}^2\hat{\mathcal{P}}_3
\end{equation} 
and Ai is the Airy function, implying
\begin{equation}
\begin{aligned}
\text{Ai}''(z) -z\text{Ai}(z) = 0,\\
\text{Ai}'''(z)-z\text{Ai}'(z)-\text{Ai}(z) = 0.
\end{aligned}
\end{equation}
We next posit the following expansions for $F$ and $G$,
\begin{equation}
\begin{aligned}
F\sim F_0(z)+\sum \left(\frac{\tilde{\Delta}}{i}\right)^mF_m(\mathcal{Y},z),\\
G\sim G_0(z)+\sum \left(\frac{\tilde{\Delta}}{i}\right)^mG_m(\mathcal{Y},z),\\
\end{aligned}
\end{equation}
and substitute the new ansatz and expansions into the governing NS equations.  

At leading order, we find that 
\begin{equation}\label{eq:CausticEikonal}
\tilde{\alpha}^2+\tilde{\xi}(\partial_z\tilde{\xi})^2= \tilde{k}^2.\\
\end{equation}
Equation (\ref{eq:CausticEikonal}) is analogous to the Eikonal equation obtained using the WKBJ ansatz (\emph{viz.} $\tilde{\alpha}^2+\tilde{\beta}^2=\tilde{k}^2$), and its solutions can be expressed as
\begin{equation}
\begin{aligned}
\frac{2}{3}\tilde{\xi}^{3/2} = -\int_{z_c}^z \left(\tilde{k}^2(t)-\tilde{\alpha}^2\right)^{1/2}dt \quad z<z_c,\\
\frac{2}{3}(-\tilde{\xi})^{3/2} = \int_{z_c}^z \left(\tilde{\alpha}^2-\tilde{k}^2(t)\right)^{1/2}dt \quad z>z_c,\\
\end{aligned}
\end{equation}
where $z_c$ is the spanwise location of the caustic.  The leading-order terms in the expansions for $F$ and $G$ are found to be
\begin{equation}
\begin{aligned}
F_0(z) = \frac{\gamma_f \tilde{\lambda}}{\hat{\mathcal{P}}_3|_{\mathcal{Y} = 0}(\partial_z\tilde{\xi})^{1/2}},\\
G_0(z) = \frac{\gamma_g \tilde{\lambda}}{\hat{\mathcal{P}}_3|_{\mathcal{Y} = 0}(\tilde{\xi}\partial_z\tilde{\xi})^{1/2}}.\\
\end{aligned}
\end{equation}
Here, the coefficients $\gamma_f$ and $\gamma_g$ are constants of integration that are determinable by matching with the WKBJ solution away from the caustic.  It can be shown that $(\tilde{k}^2-\tilde{\alpha}^2)$ has a simple zero at $z = z_c$. Consequently, in the region where $(z-z_c)$ is small, $\tilde{\xi}$ is proportional to $-(z-z_c)$. To avoid singular growth near the caustic we choose $\gamma_g = 0$. Then the leading asymptotic approximation in the region where $\tilde{\xi}(z)/\tilde{\Delta}^{2/3}=\mathit{O}(1)$ is
\begin{equation}
\mathcal{P}' \sim \frac{\tilde{\Delta}^{17/6}\gamma_f\hat{\mathcal{P}}_3\tilde{\lambda}}{\hat{\mathcal{P}}_3|_{\mathcal{Y} = 0}(\partial_z\tilde{\xi})^{1/2}} \text{Ai}\left(-\frac{\tilde{\xi}(z)}{\tilde{\Delta}^{2/3}}\right)e^{i(\tilde{\alpha}/\tilde{\Delta})(x-c t)}+c.c.
\end{equation}
In the illuminated region away from the caustic, the Airy function can be replaced by its asymptotic expansion for large negative argument to give 
\begin{eqnarray}
\mathcal{P}'&\sim&-i\frac{\tilde{\Delta}^3\gamma_f \hat{\mathcal{P}}_3\tilde{\lambda}}{2\sqrt{\pi}\hat{\mathcal{P}}_3|_{\mathcal{Y} = 0}(\tilde{k}^2-\tilde{\alpha}^2)^{1/4}} \left[ e^{i\left(\frac{1}{\tilde{\Delta}} \int_z^0(\tilde{k}^2-\tilde{\alpha}^2)^{1/2}dz+\pi/4\right)}+e^{i\left(-\frac{1}{\tilde{\Delta}} \int_z^0(\tilde{k}^2-\tilde{\alpha}^2)^{1/2}dz+3\pi/4\right)}\right]\nonumber\\
&&\qquad\qquad\qquad\qquad\qquad\times\;\;e^{i(\tilde{\alpha}/\tilde{\Delta})(x-c t)}+c.c.
\end{eqnarray}
for $z<z_c$ and $z_c>0$. This approximation shows that the ``incident'' instability mode is reflected at the caustic.  The reflected mode has the same amplitude as the incident mode but experiences a $\pi/2$ shift in relative phase upon entering the illuminated region.  In contrast, for $z>z_c$ and $z_c>0$, the asymptotic approximation of $\mathcal{P}'$ for $|\tilde{\xi}(z)/\tilde{\Delta}^{2/3}| \gg 1$ is
\begin{equation}
\mathcal{P}' \sim \frac{\tilde{\Delta}^3\gamma_f \hat{\mathcal{P}}_3\tilde{\lambda }}{2\sqrt{\pi}\hat{\mathcal{P}}_3|_{\mathcal{Y} = 0}(\tilde{\alpha}^2-\tilde{k}^2)^{1/4}} e^{-\frac{1}{\tilde{\Delta}} \int_z^0(\tilde{\alpha}^2-\tilde{k}^2)^{1/2}dz} e^{i(\tilde{\alpha}/\tilde{\Delta})(x-c t)}+c.c.,
\end{equation}
which rapidly decays to zero in the shadow zone beyond the caustic. Analogous results hold for the caustic at $z=z_c$ and $z_c<0$.

The caustics are asymptotically thin (relative to the scale of the rolls) but the amplitude of the instability mode is asymptotically large there relative to its value in the illuminated region.  Accordingly, we now estimate the contributions to the energy-budget integral on the left-hand side of (\ref{eq:KEbalance}), which subsequently is used to compute $\bar{\Omega}_c$, made by the Rayleigh mode in the caustic regions.  
To isolate these contributions, we split the integral in (\ref{eq:KEbalance}) into three parts:
\begin{eqnarray}
\int_0^{L_z/2}\tilde{w}_2|_{y=0}\partial_\mathcal{Y}\tilde{\mathcal{W}}_3|_{\mathcal{Y}=0^+}dz&=&
  \int_0^{z_c^-}\tilde{w}_2|_{y=0}\partial_\mathcal{Y}\tilde{\mathcal{W}}_3|_{\mathcal{Y}=0^+}dz\,+\,
  \int_{z_c^-}^{z_c^+}\tilde{w}_2|_{y=0}\partial_\mathcal{Y}\tilde{\mathcal{W}}_3|_{\mathcal{Y}=0^+}dz\nonumber\\
 &+&\,\int_{z_c^+}^{L_z/2}\tilde{w}_2|_{y=0}\partial_\mathcal{Y}\tilde{\mathcal{W}}_3|_{\mathcal{Y}=0^+}dz.\label{eq:CausticIntegral}
\end{eqnarray}
The amplitude $A(z)$ in the shadow region is exponentially small, so the integral from $z_c^+$ to $L_z/2$ is negligible.  Employing an analysis similar to that used within the CL away from the caustic, it can be shown that the jump condition across the CL in the neighborhood of the caustic becomes
\begin{equation}
[\partial_{\mathcal{Y}}\tilde{\mathcal{W}}_3]_-^+=\tilde{\Delta}^{1/3}\frac{2n_0\gamma_f^2}{\tilde{\alpha}^{5/3}\bar{\Omega}_c^{7/3}}\partial_z\left(\tilde{\lambda}^{1/3}\partial_z\tilde{\xi}\,\text{Ai}'^2\right).\label{eq:CausticJumpCondition}
\end{equation}
Upon using (\ref{eq:CausticJumpCondition}) to evaluate the second integral on the right-hand side of (\ref{eq:CausticIntegral}), we note that the $z$ derivative is large, $\mathit{O}(\tilde{\Delta}^{-2/3})$, since the argument of Ai is $\tilde{\Delta}^{-2/3}\tilde{\xi}(z)$. (In contrast, the notation Ai$'$ signifies an $\mathit{O}(1)$ derivative of Ai with respect to its complete argument.) This amplification, however, is offset by the small [i.e. $\mathit{O}(\tilde{\Delta}^{2/3})$] integration range.  Consequently, the second integral in (\ref{eq:CausticIntegral}) is asymptotically smaller than the first owing to the $\tilde{\Delta}^{1/3}$ factor in (\ref{eq:CausticJumpCondition}).  


\section{Scaling of fissure thickness with $Re_\tau$}\label{app:VFscaling}

At a given large value of the friction Reynolds number $Re_\tau=u_\tau h/\nu$, the fissures in wall-bounded turbulent flows are non-uniformly spaced with distance from the wall.  More precisely, observations and theoretical considerations \citep{KlewickiJFM2013} indicate that the spacing between adjacent VFs within the inertial domain, i.e. the domain extending from $y^+$=$\mathit{O}(Re_\tau^{1/2})$ to $y^+$=$\mathit{O}(Re_\tau)$, increases with distance from the wall according to a geometric progression.  Thus, the average spacing $\langle l_y^+\rangle=(Re_\tau^{1/2}Re_\tau)^{1/2}=Re_\tau^{3/4}$, where the angle brackets refer to the geometric mean.  Equivalently, $\langle l_y\rangle =Re_\tau^{-1/4}h$.  In the inertial-region SSP identified in this investigation, the fissures are uniformly separated by a distance $l_y$, and it is found that the ratio of the dimensional fissure thickness to this separation distance $\Delta_f/l_y=\mathit{O}(Re^{-1/4})$, where $Re=u_\tau l_y/\nu$.  Replacing $l_y$ with $\langle l_y\rangle$ in the preceding expressions, which is tantamount to interpreting $l_y$ in the asymptotic analysis as the geometric mean spacing of an array of non-uniformly spaced fissures in a turbulent wall flow, yields the estimate
\begin{eqnarray}
\Delta_f/h&=&\mathit{O}(Re_\tau^{-7/16}),
\end{eqnarray}
in reasonable accord with the empirically determined scaling $\Delta_f/h \sim Re_\tau^{-1/2}$; see figure~\ref{fig:FissureThickness}(a).\\ 

{\cb{
\section{ECS solution algorithm}\label{app:ECSsoln}

To compute the ECS, the following algorithm is employed.
\begin{enumerate}
\item The renormalized roll velocity field ($\tilde{v}_2(y,z)$,$\tilde{w}_2(y,z)$) is obtained by dividing (\ref{eq:Vumz})--(\ref{eq:Wumz}) by the (as yet) unknown $\bar{\Omega}_c$.
\item $\overline{u}_0(y,z)$ is obtained by numerically solving (\ref{eq:UMZ_UmeanU}) using a Fourier--Chebyshev pseudospectral scheme on the domain $0\le y\le 1$, $-L_z/2\le z\le L_z/2$ subject to the symmetry conditions $\overline{u}_0(0,z)=0$ and $\overline{u}(1,z)=1$ and periodic boundary conditions in $z$.
\item The fluctuation fields within the VF are obtained by collapsing (\ref{eq:VFfluc2dCONT})--(\ref{eq:VFfluc2dv}) into Rayleigh's equation and then numerically solving the resulting 1D differential eigenvalue problem (treating the total horizontal wavenumber $k(z)$ as the eigenvalue) on the domain $-\infty<\mathcal{Y}<\infty$ using a Chebyshev collocation method, with a $\mathcal{Y}$-coordinate mapping, for a discrete set of $z$ ranging from $-L_z/2$ to $L_z/2$.
\item The coefficients $a_1$ and $a_2$ in the amplitude equation (\ref{eq:AmpEqn}) are obtained by numerical quadrature using the expressions given in (\ref{eq:a1234}), and the fluctuation amplitude function $A(z)$ is then obtained from (\ref{eq:A}).
\item The homogenized roll vorticity $\bar{\Omega}_c$ is computed via numerical quadrature using (\ref{eq:OmegEqn}).
\item Transformations (\ref{eq:transformations})--(\ref{eq:transformations3}) are utilised to obtain the fluctuation fields in the original (i.e. non-rotated) coordinates.
\end{enumerate}
Note that the fluctuation fields within the CL need not be evaluated, but if desired, they can be reconstructed using (\ref{eq:CLwfluc}) and (\ref{eq:CLflucCONT})--(\ref{eq:CLflucU}).
}}

\bibliographystyle{jfm}
\bibliography{References}
\end{document}